\documentclass[paper]{JHEP}
\usepackage{amsmath}
\usepackage{epsfig}

\def\al{\alpha}
\def\as{\alpha_{\mbox{\scriptsize s}}}
\def\aef{\alpha_{\mbox{\scriptsize eff}}}
\def\daef{\delta \aef}
\def\qq{q\bar{q}}
\def\ee{e^+e^-}
\def\LQCD{\Lambda_{\mbox{\scriptsize QCD}}}
\def\MSbar{\overline{\mbox{MS}}}
\def\MSbar{\overline{\mbox{\scriptsize MS}}}

\def\be{\beta}
\def\bbe{{\beta_T}}

\def\eps{\epsilon}
\def\de{\delta}
\def\Om{\Omega}
\def\om{\omega}
\def\gam{\gamma}

\def\Im{{\mathop{\rm Im }}}
\def\Re{{\mathop{\rm Re }}}
\def\naive{\mbox{\scriptsize naive}}
\def\jet{\mbox{\scriptsize jet}}
\def\out{\mbox{\scriptsize out}}

\def\cO#1{{\cal{O}}\left(#1\right)}
\def\half{\mbox{\small $\frac{1}{2}$}}

\def\abs#1{\left| \: #1 \: \right|}

\def\VEV#1{\left\langle#1\right\rangle}

\def\PT{\mbox{\scriptsize PT}}
\def\NP{\mbox{\scriptsize NP}}
\def\conf{\delta}
\def\cp{\lambda^{\NP}}
\def\bmu{\bar{\mu}}
\def\bnu{\bar{\nu}}
\def\Ko{K_{\out}}
\def\bKo{\bar{K}_{\out}}
\def\ka{\kappa}
\def\vka{\vec{\ka}}                

\def\cF{{\cal{F}}}
\def\cD{{\cal{D}}}

\def\cM{{\cal{M}}}
\def\cS{{\cal{S}}}

\def\cC{{\cal{C}}}
\def\cR{{\cal{R}}}

\def\bR{{\mbox{\bf R}}}

\def\tchi{\tilde{\chi}}
\def\tf{{\tilde f}}

 \newskip\humongous \humongous=0pt plus 1000pt minus 1000pt
   \newif\ifdtup

\def\fun#1#2{\lower3.6pt\vbox{\baselineskip0pt\lineskip.9pt
  \ialign{$\mathsurround=0pt#1\hfil##\hfil$\crcr#2\crcr\sim\crcr}}}
\def\epj#1#2#3{{\it Eur. Phys. J. }{\bf C #1} (#2) #3}

\title{Non-perturbative QCD analysis of \\
  near-to-planar three-jet events}

\author{
A.~Banfi,\\ 
Dipartimento di Fisica, Universit{\`a} di Milano--Bicocca
and INFN, Sezione di Milano, Italy} 
\author{
Yu.L.~Dokshitzer,\\
LPT, Universit\'e de Paris XI, Centre d'Orsay, 
France \footnote{on leave from 
PNPI, Gatchina, St.~Petersburg,
Russia}}

\author{
G.~Marchesini,\\
Dipartimento di Fisica, Universit{\`a} di Milano--Bicocca
and INFN, Sezione di Milano, Italy} 

\author{
G.~Zanderighi.\\
Dipartimento di Fisica, Universit{\`a} di  Pavia
and INFN, Sezione di Pavia, Italy} 

\abstract{
We present the analysis of the $1/Q$--suppressed non-perturbative (NP)
contribution to the cumulative out-of-event-plane momentum
distribution in $\ee$ annihilation in the near-to-planar three-jet
region.  It complements our previous study of the perturbative (PT)
distributions resummed to single logarithmic accuracy.
Due to inter-jet soft gluon radiation, the NP contributions (as well
as the PT distributions) are sensitive to the geometry (the angles
between jets) and the colour structure of the underlying hard process
(topology of the three-prong parton antenna).
The results and techniques presented here could be extended to other
multi-jet processes and, in particular, to hadron-hadron collisions.
}

\keywords{QCD, Jets, LEP and SLC Physics, NLO Computations,  
  Nonperturbative Effects}

\preprint{
     Bicocca--FT--00/17\\
     LPT--Orsay--01--04\\
     Pavia--FNT/T-00/19\\
     hep-ph/0101205\\
     January 2001}
\begin{document}

\section{Introduction \label{sec:Int}}
Multi-jet ensembles are typical for hard hadron-hadron collisions.
Production of large-$p_t$ hadrons has mainly the topology of a
four-jet event; in $W^{\pm}, Z^0$ or large-$p_t$ photon production
there are three hard jets: two formed by the initial state radiation
and one from the hard parton recoiling against the electro-weak boson.

As a first step in improving the QCD description of processes
involving more than two jets in the final state, we study three-jet
hadronic systems in a simpler environment --- without hadrons in the
initial state.
In this paper we continue the QCD analysis \cite{acopt} of
near-to-planar three-jet events in $\ee$ annihilation into hadrons in
the regime
\begin{equation}
  \label{eq:TTM}
  T\sim T_M \gg T_m=\frac{\Ko}{Q}\>,
\end{equation}
where $T,T_M$, and $T_m$ are the thrust, the thrust major and the
thrust minor, respectively.

We study the distribution in $\Ko$ in three-jet events and aim to
reach the same theoretical accuracy that is available these days for
the description of the typical $\ee$ hadronic systems, that is two-jet
events.  The collinear and infrared safe (CIS) distributions in thrust
$T$, $C$-parameter, jet masses $M^2$ and broadenings $B$ in the region
of two narrow jets ($1\!-\!T,C,M^2/Q^2,B\ll1$) have been intensively
studied in recent years.
Both the perturbative (PT) contribution and the leading
non-perturbative (NP) corrections have been computed at high accuracy.
The PT results \cite{PTstandards} involve all-order resummation of
double- (DL) and single-logarithmic (SL) contributions and matching of
the approximate resummed expressions with the exact second order
matrix elements.
The leading NP corrections \cite{NPstandards}-\cite{DMW} have been
computed at two-loop order to take into account effects of
non-inclusiveness of jet observables. The $1/Q$--suppressed NP power
corrections are needed not only to make quantitative predictions of
event shapes but to expand our knowledge of hadronization phenomena
and of the interaction in the confinement region in general.

In \cite{acopt} we have computed the SL-resummed PT expression for the
{\em integrated}\/ distribution $\Sigma(\Ko)$ (the $\Ko$-spectrum
integrated up to a given $\Ko$) in the three-jet region
\eqref{eq:TTM}.  We considered the cumulative out-of-plane momentum
both in the {\it total}\/ phase space ($\Ko^T$) and in the restricted
{\it right}\/ region ($\Ko^R$), that is in the narrow-jet hemisphere.
The results turned out to be quite interesting and informative. 
In particular at the SL-level, the PT-distribution was found to be
sensitive both to the event geometry (the angles between jets) and the
colour structure of the underlying hard event (specific hard parton
configuration).

In this paper we study the leading NP corrections to the above PT
results.  The NP contributions originate from the part of the phase
space that corresponds to emission of secondary gluons from ``large''
distances, where the QCD coupling runs into trouble and the PT treatment
becomes questionable.  Our present understanding of QCD does not
provide a ``fundamental'' solution to the problem.
For a CIS observable, however, this part of the phase space does not
lead to divergences (neither collinear nor soft), but gives rise to a
genuine NP contribution.  The latter cannot be calculated from ``first
principles''.  The only firm information we have is that long-distance
contributions bring in corrections to the PT results that are
suppressed as an inverse power of the hard scale $Q$.  The exponent
$p$ of the suppression factor $Q^{-p}$ can be extracted from the
analysis of the divergence of the formal PT series (renormalon
analysis, for a recent review see \cite{Beneke}).

To make quantitative predictions and to make it possible to compare NP
contributions to different observables, it has been suggested
\cite{NPstandards} to assume that the running coupling can still be
defined at large distances.  This simple assumption allows one to give
meaning to PT contributions at any order and to relate the magnitudes
of the power corrections with momentum averages of the QCD coupling in
the infrared region.

We follow the dispersive method \cite{DMW} which extends the notion of
the QCD coupling to small momentum scales and helps to extract and
quantify the large-distance contribution to a given CIS observable.
We show that the leading NP correction to $\Ko$ distributions and
means belongs to the same class as the two-jet shape observables
listed above \cite{DLMS}.

The main results, both in PT and NP sectors, were announced and their
physical aspects anticipated in a short letter \cite{acoletter}.  In
this paper we present the derivation of the NP contributions to
$\Ko^{T/R}$-distributions and means.

The results for the $\Ko$-distributions that are common to the
$1\!-\!T,\,M^2/Q^2,\,C$ and $B$ distributions are:
\begin{itemize}
\item the leading NP corrections are of order $1/Q$.  Since the actual
  expansion parameter is $1/\Ko$ rather than $1/Q$, they are leading
  as long as $\Ko$ is larger than $\LQCD$. For $\Ko$ of the order of
  $\LQCD$, higher powers become equally important \cite{KS};
\item the leading NP corrections in the region $\LQCD\ll\Ko\ll Q$ can
  be embodied in a {\em shift}\/ $\de\Ko$ of the argument of the PT
  distributions;
\item the shift is a logarithmic function of $\Ko$ and is determined
  analytically by perturbative calculations, except for the overall
  normalization parameter $\cp$;
\item the phenomenological NP parameter $\cp$ is given by the momentum
  integral of the QCD coupling in the infrared region. It is the same
  parameter that determines NP corrections to the two-jet shape
  observables ($T,M^2,C,B$).
\end{itemize}
Specific features of the $\Ko^{T/R}$-distributions are the following:
\begin{itemize}
\item for the {\it right}\/ $\Ko^R$-distribution, the shift behaves
  logarithmically as $\ln\frac{Q^{\NP}}{\Ko}$, which behaviour is
  similar to that of the single-jet broadening \cite{broad};
\item for the {\it total}\/ $\Ko^T$-distribution, the logarithmic
  contribution $\ln\frac{Q^{\NP}}{\Ko}$ is supplemented with a
  singular constant term proportional to $1/\sqrt{\as}$ which
  contribution has a weird colour structure and dominates at
  moderately small $\Ko$.
  This behaviour is due to an interplay between PT and NP effects and
  is reminiscent of the total broadening case \cite{broad};
\item the relevant hard scales $Q^{\NP}$ are affected both by the
  small- and large-angle radiation of secondary partons. For the
  $\Ko^T$ distribution, these scales have a simple geometrical
  interpretation.  As a result, the NP corrections, as well as the PT
  distributions, depend on geometry and colour structure of the hard
  event.
\end{itemize}
A certain similarity between the $\Ko$ and the broadening distribution
is due to the fact that both these observables accumulate
contributions from arbitrarily large soft parton rapidities (while
$T,M^2,C$ are dominated by finite rapidities --- large-angle radiation
only). As found in \cite{broad}, it is this feature that is
responsible for the logarithmic behaviour and the singular
$1/\sqrt{\as}$ contributions to the shift.

The need to present reproduceable derivations can sometimes be
detrimental to readability of a paper. This is especially true for
the present topic where the tricky event-plane kinematics and the NLO
accuracy lead to technical complications.  We have nevertheless tried
our best to keep the paper readable.

In section \ref{sec:Distribution} we set up the kinematics, recall the
PT results that will be needed for the NP analysis, and describe the
NP corrections to the Sudakov exponents (radiators).
Sections \ref{sec:T-NP} and \ref{sec:R-NP} contain the new results of
this study, namely the NP corrections to the distributions and means
of $\Ko^T$ and $\Ko^R$ observables.
In section \ref{sec:Discuss} we summarize the results 
and discuss their physical properties.
The detailed calculations are presented in a series of Appendices.

\section{Distributions and resummation\label{sec:Distribution}}
In the near-to-planar three-jet region \eqref{eq:TTM} the event, at
parton level, can be treated as a hard quark-antiquark-gluon system
accompanied by an ensemble of secondary partons $k_i$. We denote by
$p_a$ ($a=1,2,3$) the three energy-ordered hard parton momenta:
$p_1>p_2>p_3$.

Defining the event plane as the $\{yz\}$-plane (we set the thrust axis
and the thrust-major axis equal to the $z$- and $y$-axis,
respectively), the {\it total}\/ out-of-plane momentum variable is
defined by
\begin{equation}
  \label{eq:KodefT}
\Ko^T\>=\>\sum_{a=1}^3\abs{p_{ax}}+\sum_{i=1}^n\abs{k_{ix}}\,.
\end{equation}
We consider also the observable restricted to the {\it right}\/ region ($R$),
that is the hemisphere with the lowest broadening: 
\begin{equation}
  \label{eq:KodefR}
\Ko^R\>=\>\abs{p_{1x}}+\sum_{i\in R}\abs{k_{ix}}\,.
\end{equation}
The hardest parton $p_1$ lies in the right hemisphere, near the
thrust axis.

In this section we first set up the kinematics of partons in the
region \eqref{eq:TTM} and define the event plane. We then recall the
PT results that are needed for the NP analysis, in particular the
factorization of soft emission and the result of SL resummation
(details are reported in Appendix \ref{App:RadPT}). We finally report
and discuss the NP corrections to the Sudakov exponents (radiators)
that are evaluated in Appendix \ref{App:RadNP}.

\subsection{Parton kinematics and the observable}
At Born level one considers only the hard $\qq,g$ system and neglects
the soft radiation so that $\Ko=0$. We denote the three hard massless
Born momenta by $P_a$, take $P_1$ along the thrust axis and $P_2$ more
energetic than $P_3$, see Fig.~\ref{fig:Born}.  Given $T$ and $T_M$ we
have\footnote{To define the massless momenta $P_a$ one requires that
  $T$ and $T_M$ are restricted to the region
\begin{equation*}
  \label{eq:kinB}
\frac{2(1-T)\sqrt{2T-1}}{T}<T_M<\sqrt{1-T}\>.
\end{equation*}
In the following we consider $T,T_M$ restricted to this region.} 
\begin{equation}
  \label{eq:Pa}
\begin{split}
&  P_1=E(T,0,0,T)\>,\quad
   P_2=E(x_2,0, T_M,-t_{2})\>,\quad
   P_3=E(x_3,0,-T_M,-t_{3})\>, \\
&  x_2=\frac{2-T}{2} +\frac{T}{2}\rho\>,\qquad 
   t_2=\frac{T}{2} +\frac{2-T}{2}\rho\>,\qquad
   \rho\equiv\sqrt{1-\frac{T_M^2}{1-T}}\>,\quad E=\frac{Q}{2}\>.
  \end{split}
\end{equation}
The fractions $x_3$ and $t_3$ are obtained from $2=T+x_2+x_3$ and
$T=t_2+t_3$.

There are essentially three underlying configurations for the hard
partons according to which is the parton momentum corresponding to the
gluon. We denote by $\conf$ the configuration in which the momentum of
the hard gluon is $P_{\conf}$.  In Fig.~\ref{fig:Born} we represent
the three Born configurations.

\EPSFIGURE[ht]{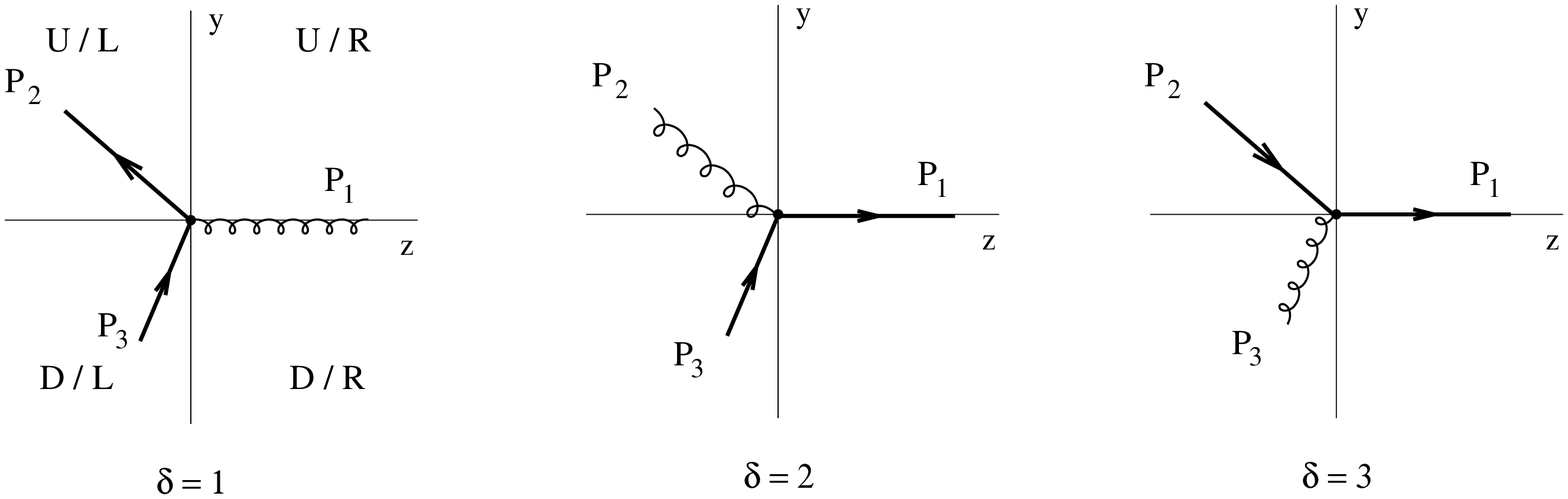,width=1.0\textwidth}{The three Born
  configurations $\conf$ for $T=0.75$ and $T_M=0.48$ ordered according
  to increasing probability.  The thrust $T$ and thrust major $T_M$
  are along the $z$- and $y$-axis respectively. The up, down, left and
  right hemispheres ($U,D,L,R$ respectively) are
  indicated.\label{fig:Born}}

Beyond the Born level, the three hard parton momenta $p_a$ no longer
lie in the plane but are displaced from $P_a$ by ``soft recoils''
$q_a$:
\begin{equation}
  \label{eq:Pa'}
p_a=P_a+q_a\>.  
\end{equation}
As we have shown in \cite{acopt}, finite rescaling of the in-plane
momenta $p_{az}$, $p_{2y},p_{3y}$, due to hard collinear splittings,
gets absorbed into the first hard correction to the emission
probability of soft gluons, which is then resummed and embodied into
the radiator.  Bearing this in mind, all recoils $q_a$ and the
secondary parton momenta $k_i$ can be treated as small (soft).

Due to the fact that $p_{az}\sim P_{az}\!=\!\cO{Q}$, $p_{2y}\sim
P_{2y}\!=\!\cO{Q}$ and $p_{3y}\sim P_{3y}\!=\!\cO{Q}$, the five soft
{\em in-plane}\/ recoil components $q_{az},q_{2y},q_{3y}$ can be
neglected both in the definition of the event plane and in the soft
matrix elements.  Then, the definition of the event plane involves the
remaining four recoil components and results in the following
kinematical constraints:
\begin{equation}
  \label{eq:plane}
\begin{split}
  &\vec{q}_{1t}+ \sum_R \vec{k}_{it}=0\>,
\qquad \vec{q}_{1t}\equiv (q_{1x},q_{1y})\>,
\qquad \vec{k}_{it}\equiv (k_{ix},k_{iy})
\\& q_{2x}+q^+_{1x}+\sum_U k_{ix}=0\>, \quad
  q_{3x}+q^-_{1x}+\sum_D k_{ix}=0\>, \qquad q^{\pm}_{1x} \equiv
  q_{1x}\> \vartheta(\pm q_{1y})\>,
\end{split}
\end{equation}
where we denote by $R,\,L,\,U$ and $D$ the right-, left-, up- and down-
phase space regions, see Fig.~\ref{fig:Born}.
The total and the right-hemisphere $\Ko$ become
\begin{equation}
  \label{eq:Ko}
\begin{split}
& \Ko^T=|q_{1x}|+|q_{2x}|+|q_{3x}|+\sum_i|k_{ix}|\>,
\qquad 
\Ko^R=|q_{1x}|+\sum_{i\in R}|k_{ix}|\>.
\end{split}
\end{equation}

\subsection{Resummed distribution for $\Ko\ll Q$}
We study the integrated $\Ko$-distribution defined as
\begin{equation}
\label{eq:sigdef}
\begin{split}
\frac{d\sigma^T(\Ko)}{dTdT_{M}}\!=\!Q^5
\sum_m\!\int\! {d\sigma_m}
\delta^{3}\!\left(\sum_{h\in R}\vec{p}_{h}\!-\! \vec{P}_1\!\right)
\!\delta^{2}\!\left(\sum_{h\in U}\vec{p}_{ht}\!-\!\vec{P}_{2t}\!\right)
\!\vartheta\!\left(\!\Ko\!-\!\sum_{h=1}^{m} |p_{hx}|\!\right),
\end{split}
\end{equation}
where $m$ denotes the number of final particles with momenta $p_h$.
The two delta-functions fix the event plane and the theta-function
defines the observable.  For the total-$\Ko$ case ($\Ko^T$) the sum
runs over all particles in the event.
Analogously we define the {\em right}\/ distribution $\Ko^R$ by
restricting the sum over particles in the theta-function of
\eqref{eq:sigdef} to those belonging to the right hemisphere.

In the near-to-planar ``three-jet region'' \eqref{eq:TTM}, the
distributions in $\Ko^T$ or in $\Ko^R$ are factorized in the form
\begin{equation}
  \begin{split}
\label{eq:Sig-factorized}
\frac{d\sigma^{T/R}(\Ko)}{dTdT_{M}} =
c(\as)
\cdot \sum_{\conf=1}^3 \frac{d\sigma_{\conf}^{(0)}}{dT dT_M} 
\cdot \Sigma^{T/R}_{\conf}(\Ko)\>.
\end{split}
\end{equation}
The factor ${d\sigma_{\conf}^{(0)}}/{dT dT_M}$ is the three-jet
differential Born cross section in the configuration $\conf$.  The
factor $\Sigma^{T/R}_{\conf}$ accounts for the soft radiation emitted
by the hard $\qq,g$ system. The first factor
$$ 
    c(\as) = 1+\cO{\as} 
$$
is the non-logarithmic coefficient function which takes into account
corrections due to hard large-angle radiation not included in
$\Sigma^{T/R}_{\conf}(\Ko)$.  The distribution $\Sigma^{T/R}_{\conf}$
is given by
\begin{equation}
\label{eq:Sigdef}
\begin{split}
\Sigma^{T/R}_{\conf}(\Ko)\>&=\>
\sum_n \frac1{n!} \int \>dH^{T/R}_n(\Ko)\> 
\prod_{i}^n \frac{d^3k_i}{\pi \om_i}\> M^2_{n,\,\conf}\>.
\end{split}
\end{equation}
Here $M^2_{n,\,\conf}$ is the distribution for the emission of $n$
soft partons from the primary $\qq,g$ system in the configuration
$\conf$ and $dH^{T/R}_n(\Ko)$ is (a part of) the phase space,
involving the three-jet kinematics \eqref{eq:plane} and the observable
\eqref{eq:Ko}.

The procedure used to resum the PT contributions of the
$\Ko$-distribution is described in detail in \cite{acopt} and recalled
in Appendix~\ref{App:RadPT}.  The essential point which allows
resummation of $\Sigma^{T/R}_{\conf}(\Ko)$ is the factorization of
the soft distribution $M_{n,\conf}^2$ and of the phase space
$dH^{T/R}_n(\Ko)$. To achieve SL accuracy one needs to consider the
soft emission factorization of $M_{n,\conf}^2$ at the two-loop level.

Resummation employs the Mellin--Fourier representation of
$\Sigma^{T/R}_{\conf}(\Ko)$ and gives rise to the exponential of the
radiator. To describe this structure (see \cite{acopt}, Appendices
\ref{App:RadPT} and \ref{App:RadNP}), we start from the phase space
factor $dH^{T}_n$ which is given by
\begin{equation}
\label{eq:dHnT}
\begin{split}
&dH^T_n(\Ko)\>\equiv\>d^2\vec{q}_{1t} dq_{2x} dq_{3x}\>
\vartheta\left(\!\Ko\!-\!\sum_{a=1}^3\abs{q_{ax}}\!-\!
\sum_{i=1}^n\abs{k_{ix}}\!\right)\\
&\times 
  \delta^2\!\left(\vec{q}_{1t}\!+\!  \sum_R \vec{k}_{it}\right)
  \delta\!\left(q_{2x}\!+\!q^+_{1x}\!+\!\sum_U k_{ix}\right)
  \delta\!\left(q_{3x}\!+\!q^-_{1x}\!+\!\sum_D k_{ix}\right),
\end{split}
\end{equation}
where the delta functions define the event plane \eqref{eq:plane}.
Correspondingly, the phase space factor $dH^{R}_n$ involves only
partons emitted in the right region:
\begin{equation}
\label{eq:dHnR}
\begin{split}
dH^R_n(\Ko)\>&\equiv\>dq_{1x}\>
\vartheta\left(\!\Ko\!-\!\abs{q_{1x}}\!-\!\sum_{R}\abs{k_{ix}}\!\right)  
\delta\!\left(q_{1x}\!+\!\sum_R k_{ix}\right).
\end{split}
\end{equation}
In the latter case, since the distribution is insensitive to the soft
recoil components $q_{2x},q_{3x}$ and $q_{1y}$, they can be neglected
in the matrix element and integrated over.

To factorize the multi-parton emissions in the phase space factors
$dH^{T/R}_n$ we use the Mellin and Fourier representations for the
theta- and delta-functions in \eqref{eq:dHnT} and \eqref{eq:dHnR}.
This results in
\begin{equation}
  \label{eq:dhn}
  dH^{T/R}_n=\int\frac{d\nu\,e^{\nu\Ko}}{2\pi i \nu}\>dh^{T/R}_n\>,
\qquad
  dh^{T/R}_n=\int d\mu^{T/R}\>\prod_{i=1}^n u^{T/R}(k_i)\>,
\end{equation}
where the integration elements $d\mu^{T/R}$ and the parton {\em
  sources}\/ $u^{T/R}(k_i)$ are described in Appendix~\ref{App:RadPT}
and \ref{App:RadNP}. Here we report some essential points concerning
the integration and the sources.

\paragraph {Integration.} 
For the $\Ko^T$ case we need to introduce four Fourier variables
$\gam,\be_a$ conjugated with $q_{1y}$ and $q_{ax}$, respectively.  For
the $\Ko^R$ case only one Fourier variable $\be$ conjugated with
$q_{1x}$ is needed.  The two integrations are
\begin{equation}
  \label{eq:dmu}
\begin{split}
&\int d\mu^T\equiv \frac{\nu^4}{8}
\int_{-\infty}^{\infty}\frac{d\gam dq_{1y}}{2\pi} 
\prod_{a=1}^3\int_{-\infty}^{\infty}\frac{d\be_a dq_{ax}}{\pi}
\>V^T(\gam,\be, q)\>,\\
&\int d\mu^R\equiv \frac{\nu}{2}
\int_{-\infty}^{\infty}\frac{d\be dq_{1x}}{\pi} 
\>V^R(\be, q_{1x})\>.
\end{split}
\end{equation}
Here we have rescaled the Fourier variables by $\nu$, so that they are
dimensionless. The functions $V^{T/R}$ are given in \eqref{eq:VT} and
\eqref{eq:VR}. They are the Mellin--Fourier factors involving the
recoil momenta.

\paragraph {Sources.} 
The soft-parton sources $u^{T/R}(k_i)$ contain Mellin -- Fourier
factors involving the secondary parton momenta.  The sources enter as
the factors $[u(k_i)\!-\!1]$ integrated with the radiation probability
$dw(k_i)$, where the subtraction unity corresponds to the virtual
contribution.  The source function tends to unity, $u(k_i)\!\to\!1$,
when $k_i$ becomes collinear to one of the radiating hard partons
$p_a$. This ensures cancellation of the singularities in the soft
matrix element $M^2_{n,\conf}$.

As has been already mentioned, for the sake of the PT-resummation the
parton recoil effects can be neglected in the soft matrix element
\cite{broad}, so that we can take $u(k_i)\!\to\! 1$ for $k_i$
collinear to the skeleton-momenta $P_a$ (see \eqref{eq:u0ell} and
\eqref{eq:u0}).
In the calculation of the NP corrections, however, it is essential to
keep the recoil in the soft matrix element, so as described in detail
in Appendix~\ref{App:RadNP}.  This makes it necessary to accurately
define the sources with account of the recoil as well, as to preserve
the real-virtual cancellations in the collinear limit, see
\eqref{eq:uabT} and \eqref{eq:u1bR}.

The $\Ko$-distributions share this feature with the broadenings
\cite{broad}.  In both cases the observables are uniform in secondary
parton rapidities. In our case, accumulation of the NP contribution to
$\Ko$ over large rapidities is bounded by the mismatch between the
actual parton momentum $p_a$ and the corresponding ``Born'' momentum
$P_a$ (which lies in the event plane).  Thus, the rapidity integrals
turn out to be essentially the logarithm of the angle between the hard
parton $\#a$ and the event plane.

From \eqref{eq:dhn} we obtain the Mellin representation of
$\Sigma^{T/R}_{\conf}(\Ko)$:
\begin{equation}
  \label{eq:mellin}
  \Sigma^{T/R}_{\conf}(\Ko)=\int\frac{d\nu\,e^{\nu\Ko}}{2\pi i \nu}
\>\sigma^{T/R}_{\conf}(\nu)\>,
\end{equation}
where the contour runs parallel to the imaginary axis with $\Re\nu>0$.
The Mellin moment $\sigma^{T/R}_{\conf}(\nu)$ is given by the sum, see
\eqref{eq:Sigdef},
\begin{equation}
  \label{eq:sigmapiccolo}
\begin{split}
&  \sigma^{T/R}_{\conf}(\nu)\>
=\>\sum_n \frac1{n!} \int \>dh^{T/R}_n(\Ko)\> 
\prod_{i}^n \frac{d^3k_i}{\pi \om_i}\> M^2_{n,\,\conf}\\
&=\>\int d\mu^{T/R} 
\left\{\sum_n \frac1{n!} \int \prod_{i}^n \frac{d^3k_i}{\pi\om_i}\>
u^{T/R}(k_i)\cdot M^2_{n,\,\conf}\right\}
\>=\> \int d\mu^{T/R}\>e^{-\cR^{T/R}}\>,
\end{split}
\end{equation}
where the radiator $\cR^{T/R}$ at SL level is obtained by using soft
emission factorization of $M_{n,\conf}^2$ at two loops.  The radiator
depends on the hard momenta $P_a$, the Mellin and Fourier variables
and, in general, on the recoil momenta.

The Mellin moment $\sigma^{T/R}_{\conf}(\nu)$ is a regular function of
$\nu$ in the entire complex plane. It decreases for $\Re\nu\to \infty$
so that
\begin{equation}
  \label{eq:bound1}
  \Sigma^{T/R}_{\conf}(\Ko<0)=0\>.
\end{equation}
The fact that $\sigma^{T/R}_{\conf}(\nu)\propto e^{\nu\Ko^m}$ for  
$\Re\nu\to -\infty$, 
where $\Ko^m$ is the maximum kinematical value of $\Ko^{T/R}$,   
ensures that 
\begin{equation}
  \label{eq:bound2}
  \Sigma^{T/R}_{\conf}(\Ko>\Ko^m)=\sigma_{\conf}^{T/R}(0)=1\>.
\end{equation}
The near-to-planar kinematics $\Ko\ll Q$ corresponds to the region of
the Mellin variable $Q\nu\gg1$.

\subsection{NP correction to the mean}
From the Mellin representation \eqref{eq:mellin}, we formally get
\begin{equation}
  \label{eq:mean}
  \VEV{\Ko^{T/R}}_{\conf}\equiv\int_0^{\infty}KdK\>
\frac{d\Sigma_{\conf}^{T/R}(K)}{dK}\>=\>
-\partial_{\nu}\sigma_{\conf}^{T/R}(\nu)|_{\nu=0}\>,
\end{equation}
where, due to \eqref{eq:bound2}, the integration over $K$ has been
extended to infinity.  This quantity, at the PT level, is determined
by the region $\Ko\sim Q$ where the soft approximation is invalid.
However, the NP correction originates, in any case, from the region of
small (transverse) momenta of the order of $\LQCD$.
Therefore the NP correction to the mean $\Ko^{T/R}$ can be computed
within the present formalism based on the soft factorization.  The NP
correction to the mean is then obtained, according to \eqref{eq:mean},
from the NP contribution to $\sigma_{\conf}^{T/R}(\nu)$ at $\nu=0$.
Even though $Q\nu=\cO{1}$, the NP answer is still dominated by small
secondary gluon momenta, that is, by large values of the Fourier
variables $\be_a$ and $\gam$: $Q\nu\be_a\sim Q\nu\gam\gg 1$.

\subsection{$\Ko^T$ case: NP radiator and recollection of PT results}
We discuss the structure of the PT and NP components of the radiator
for the $\Ko^T$ distribution.  The detailed calculation of the NP
corrections are given in Appendix~\ref{App:RadNP}. We also recall the
essential points of the PT results both for the radiator and the
distribution (see \cite{acopt} and Appendix \ref{App:RadPT}), which
are needed to derive the NP corrections in the next section.  In this
subsection we suppress the index $T$ and the index $\conf$ marking the
jet configuration.

\subsubsection{$\Ko^T$-radiator}
The radiator for the $\Ko^T$ distribution has PT and NP components:
\begin{equation}
  \label{eq:T-Rad}
  \cR(\nu,\be,\gam,q)=
\cR^{\PT}(\nu,\be,\gam)+\de\cR(q)\>.
\end{equation}
The PT contribution depends on the Fourier variables but is
independent of the recoil momenta. (As mentioned, recoil effects at PT
level produce corrections that are beyond the SL accuracy, see
\cite{broad}).  On the contrary, the NP correction depends on the
out-of-plane $x$-components of the recoil momenta but does not depend
on $\gam,\be_a$.

We first address the NP contribution to the radiator calculated in
detail in Appendix~\ref{App:RadNP}. We have
\begin{equation}
  \label{eq:T-RadNP}
  \delta \cR(q)\>=\> 
\nu\>\cp\> \sum_a C_a\ln\frac{Q^{\NP}_{a}}{|q_{ax}|}\,.
\end{equation}
We comment on this expression.
\begin{itemize}
\item $C_a$ is the colour charge of parton $\#a$ in the configuration
  $\conf$ (here implicit). For instance, for the most probable
  configuration $\conf=3$ in which the gluon is the least energetic of
  the three hard partons, we have $C_1\!=\!C_2\!=\!C_F$ and
  $C_3\!=\!N_c$.
\item Expression \eqref{eq:T-RadNP} is obtained by linearizing the
  sources $u^T(k_i)$.  The NP parameter $\cp$ defined in
  \eqref{eq:cp}, has dimension of mass and is responsible for $1/Q$
  power corrections to jet observables.
\item Linearization makes $\de\cR$ proportional to the Mellin variable
  $\nu$ which, in turn, implies that the NP correction emerges
  essentially as a shift in $\Ko$.
\item The logarithmic behaviour in $|q_{ax}|$ comes from the soft
  gluon rapidity integration which is bounded by the rapidity of the
  parton $p_a$, with rapidities defined with respect to the
  corresponding Born momenta $P_a$.  (Radiation at smaller angles
  corresponds to collinear splitting and does not contribute to the
  observable due to real-virtual cancellation.)
\item The hard NP scale $Q^{\NP}_{a}\sim{Q}$ associated with hard
  parton $\#a$ contains two factors:
\begin{equation}
  \label{eq:QNP}
  Q^{\NP}_{a}=Q_{a}\cdot \zeta^{\NP}\>, 
\qquad \zeta^{\NP}=2\,e^{-2} 
\end{equation}
The rescaling factor $\zeta^{\NP}$ is defined in \eqref{eq:zeta} and
originates as a constant leftover after the real-virtual cancellation
of collinear singularities in the NP radiation. 

The first factor $Q_a$ is determined by large-angle soft emissions and
is sensitive to the geometry of the event as it depends on the angles
between jets.  In the configuration $\conf$ one has
\begin{equation}
  \label{eq:Qa}
  Q_{b}^2 =Q_{c}^2= 2(P_bP_c)\>, \quad
  Q_{\conf}^2 = \frac{2(P_bP_{\conf})(P_{\conf}P_c)}{(P_bP_c)}\>,
\end{equation}
where $b,c\ne\conf$ denote the quark-antiquark indices. The scales for
the fermions are equal to the invariant mass of the quark-antiquark
dipole. For the gluon, $Q_{\conf}$ is the invariant gluon transverse
momentum with respect to the $\qq$ pair.  (Notice that when the gluon
becomes collinear to the quark or the antiquark, the scale $Q_{\conf}$
decreases and the non-Abelian contribution reduces.)
\item The NP radiator does not depend on the Fourier variables
  $\be_a,\gam$.  As in the case of broadening, see \cite{broad}, this
  is due to the fact that the contributions linear in $k_x$ or $k_y$
  vanish, and the $\be_a,\gam$--dependence of the radiator emerges
  only at the $1/Q^2$ level.
\item Since the NP radiator depends on $q_{ax}$, the way it affects
  the $\Ko$-distribution depends on the PT distributions in parton
  recoils.
\end{itemize}
The NP corrections to the integrated distribution $\Sigma(\Ko)$ that
correspond to the NP radiator $\de\cR$ are analyzed in the next
section.

Now we recall the structure of the PT radiator in \eqref{eq:T-Rad}
(for details see Appendix \ref{App:RadPT}):
\begin{equation}
\label{eq:T-RadPT}
\begin{split}  
&\cR^{\PT}(\nu,\be,\gam)=
\bR_1(\bnu,\be_{12},\be_{13},\gam)
+ R_2(\rho_2)+R_3(\rho_3)\>,\\
&
\bR_1(\bnu,\be_{12},\be_{13},\gam)\equiv
\int_0^{\infty}\!\frac{dy}{\pi(1\!+\!y^2)}\!
\left\{R_1(\rho_{12})+R_1(\rho_{13})\right\}
\end{split}
\end{equation}
with ($\be_{1a}=\be_1+\be_a$)
\begin{equation}
  \label{eq:rhos}
\begin{split}
\rho_{12}&\equiv \bnu\sqrt{(1\!-\!i\gam y)^2\!+\!\be^2_{12}}\,, \quad
\rho_{13}\equiv \bnu\sqrt{(1\!+\!i\gam y)^2\!+\!\be^2_{13}}\,,\\
\rho_{a} &\equiv \bnu\sqrt{1+\be_a^2}\>,\quad a=2,3\>,\qquad
\bnu=e^{\gam_E}\nu\>.
\end{split}
\end{equation}
The three terms of $\cR^{\PT}$ in \eqref{eq:T-RadPT} describe the
radiation from partons \#1, \#2 and \#3, respectively.  The two pieces
of $\bR_1$ correspond to secondary gluon emission in the Up
($k_{y}>0$) and Down hemispheres ($k_{y}<0$).
The function $R_{a}(\rho)$ is the two-loop single-jet
radiator given by
\begin{equation}
\label{eq:r}
R_{a}(\rho)\>\equiv\> C_{a}\,r(\rho,\,Q^{\PT}_{a})\>,
\qquad
 r\left(\rho,Q^{\PT}_{a}\right)=\frac{2}{\pi}\int_{1/\rho}^{Q^{\PT}_a}
\frac{dk_x}{k_x}\as(2k_x)\ln \frac{Q^{\PT}_{a}}{k_x}\>,
\end{equation}
with the running coupling in the physical (``bremsstrahlung''; CMW)
scheme~\cite{CMW}.  (The factor $2$ in the argument of the running
coupling originates after the integration over the in-plane component
of the transverse momentum.)  The {\em perturbative}\/ hard scales are
again composed of two factors:
\begin{equation}
  \label{eq:QPT}
  Q_a^{\PT}=Q_a\cdot \zeta_a^{\PT}\,,\quad 
\zeta_a^{\PT}\!=\!\half e^{-3/4} \> \mbox{for} 
\,\> a\neq\conf \> (q\>\mbox{or}\>\bar q\,),
\quad \zeta_a^{\PT}\!=\!\half e^{-\be_0/4N_c} 
\,\> \mbox{for} \> a=\conf\> (\,g\,).
\end{equation}
The first are the same geometry-dependent factors \eqref{eq:Qa} as in
the NP scales, that account for soft large-angle emission.  The second
factors are hard SL corrections due to collinear quark ($-3C_F$) and
gluon ($-\be_0$) splittings. (The factor $1/2$ in $\zeta_a^{\PT}$,
again, comes from the integration over the in-plane components.)

The function $r$ contains both DL ($\as^nL^{n+1}$) and SL
($\as^nL^{n}$) contributions, with $L=\ln\rho Q$ and $n\ge1$.  The
precise value of the hard scale and, in particular, the geometry of
the event, is relevant only at SL level.

The same is true for the dependence of the functions $\rho$ in
\eqref{eq:rhos} on the Fourier variables $\be_a,\gam$, {\em
  provided}\/ the latter are {\em finite}.  This is the situation one
encounters when calculating the PT distribution. Here the
$\be_a,\gam$--integrals are well convergent so that
$\be_a,\gam=\cO{1}$ and we can simplify the $\be_a,\gam$--dependence
by expanding the logarithmic functions $R_a$ in \eqref{eq:T-RadPT} to
SL accuracy as
\begin{equation}
  \label{eq:SLr}
r(\rho,Q_a^{\PT}) 
=  r\left(\bnu,Q_a^{\PT}\right)+\ln\left(\frac{\rho}{\bnu}\right)
\>r'(\nu,Q_a^{\PT})\>+\> \cO{\as},
\end{equation}
where
\begin{equation}
  \label{eq:r'}
r'(\nu,Q)=\nu\partial_{\nu}\>r(\nu,Q)=
  \frac{2\as(1/\nu)}{\pi}\>\ln(\nu Q)\>.
\end{equation}
Since $r'$ is a SL function, we do not need to care about the precise
expression for the hard scale in \eqref{eq:r'} and about the
difference between $\nu$, $\bnu$, $\nu/2$ or alike.

Physically, the dominant integration region $\be_a,\gam=\cO{1}$
corresponds to the situation when the hard parton recoils and the
secondary parton momenta are of the same order:
$$
  \abs{q_{ax}}\sim \abs{k_{ix}} \sim \Ko\,.
$$
Using 
\eqref{eq:SLr} we derive then, with SL accuracy,
\begin{equation}
  \label{eq:SL-Rad}
  \cR^{\PT}(\nu,\be,\gam)\simeq \sum_aR_a(\bnu)+S(\nu,\be,\gam)\>.
\end{equation}
where $S$ is the SL function 
\begin{equation}
  \label{eq:S}
\begin{split}
S=R_1'(\nu)\!\int_0^{\infty}\!\frac{dy}{\pi(1\!+\!y^2)}\,
&\ln\sqrt{\! 
\left((1\!-\!i\gam y)^2\!+\!\be^2_{12}\right)
\left((1\!+\!i\gam y)^2\!+\!\be^2_{13}\right)}\\
&+\sum_{a=2,3}R_a'(\nu)\ln\sqrt{1\!+\!\be_a^2}\,,
\end{split}
\end{equation}
and $R'_a(\nu)=C_a\,r'(\nu,Q)$.  In the expression \eqref{eq:SL-Rad},
the geometry of the event enters through the DL radiators $R_a(\bnu)$
while the SL function $S$ contains the Fourier variable dependence and
embodies hard parton recoil effects.  Through the colour factors in
\eqref{eq:S}, the recoil effects depend on the specific configuration
$\conf$.
 
Let us stress that the approximate expression \eqref{eq:SL-Rad} can be
used only if the $\be_a,\gam$--integrals converge, which is the case
of the PT distribution.  In the evaluation of the NP correction
instead, one encounters poorly convergent (logarithmic)
$\be_a,\gam$--integrals and we cannot use the approximation
\eqref{eq:SL-Rad} but have to employ the full expression
\eqref{eq:T-RadPT}.

\subsubsection{Recollection of the PT distribution in $\Ko^T$ 
\label{sec:recollationT}}
Since the PT radiator is recoil-insensitive, we can freely integrate
over $q_{1y}$ and $q_{ax}$.  The integrals in the Fourier variables
are convergent, so that we can use the approximation \eqref{eq:SL-Rad}
and obtain, to SL accuracy, the final result
\begin{equation}
  \label{eq:T-sigPT}
  \sigma^{\PT}(\nu)\!=\!\int d\mu^T\,e^{-\cR^{\PT}(\nu,\gam,\be)}
=e^{-\sum R_a(\bnu)}\,\cF_T(\nu)\,,
\quad \cF_T(\nu)\!=\!\int d\mu^T\,e^{-S(\nu,\gam,\be)}\,, 
\end{equation}
where $d\mu^T$ is the integration introduced in
\eqref{eq:dmu}.  Here $\cF_T(\nu)$ depends on $\nu$ via the
SL function $r'(\nu,Q)$ and has been computed in \cite{acopt}. To first order
in $r'$ one has
\begin{equation}
  \label{eq:cFTone}
  \cF_T(\nu)=1-\ln2\,r'\,\left(\,2C_1+C_2+C_3\,\right)+\cO{{r'}^2}
\quad R'_a=C_a\,r'(\nu,Q)\>. 
\end{equation}
We now recall how to compute the distribution, i.e. how to perform
the $\nu$--integral in the Mellin representation \eqref{eq:mellin}. To
this end we use the identity
\begin{equation}
  \label{eq:identity}
\begin{split}  
\Sigma^{\PT} (\Ko)&=\int\frac{d\nu\,e^{\nu\Ko}}{2\pi i \nu}\,
\left\{e^{-\sum R_a(\bnu)}\cdot \cF_T(\nu)\right\}
\\&
=e^{-\sum R_a(\bKo^{-1} e^{-\partial_z})}
\cdot \cF_T(\Ko^{-1} e^{-\partial_z})
\left. \frac{1}{\Gamma(1+z)}\right|_{z=0},
\qquad \bKo=e^{-\gam_E}\Ko\>.
\end{split}
\end{equation}
The action of this differential operator on the function
$\Gamma^{-1}(1+z)$ reduces, within the adopted accuracy, simply to
replacing $z$ with $R'_T$.  The PT distribution to SL accuracy takes
the form
\begin{equation}
  \label{eq:T-SigPT}
\begin{split}
  \Sigma^{\PT} (\Ko)\>\simeq
\>e^{-\sum_aR_a\left(\bKo^{-1}\right)}\cdot
\frac{\cF_T(\Ko^{-1})}{\Gamma(1+R'_T)}\>, 
\qquad R'_T\equiv C_T\,r'\left(\Ko^{-1},Q\right)\>,
\end{split}
\end{equation}
with $C_T=2C_F+N_c$ the total colour charge.  Dependence on the
geometry enters through the hard scales in the DL radiators
$R_a(\bKo^{-1})$, while the hard parton recoils build up the SL
correction factor $\cF_T(\Ko^{-1})$.

The r\^ole of recoil effects and the event-plane kinematics can be
seen already in the first-order expansion in $\as$.  Using the
expansion of $\cF_T$ given in \eqref{eq:cFTone}, for the differential
distribution we obtain
\begin{equation}
\label{eq:T-dSigma}
  \frac{d\Sigma^{\PT}(\Ko)}{d\ln \Ko}=\frac{2\as}{\pi}\left\{
C_{1}\ln\frac{4Q_{1}^{\PT}}{\Ko}+
C_{2}\ln\frac{2Q_{2}^{\PT}}{\Ko}+
C_{3}\ln\frac{2Q_{3}^{\PT}}{\Ko}\right\}+\>\cO{\as^2}\,.
\end{equation}
Gluon emission off the parton $p_a$ has its characteristic hard scale
$Q_a^{\PT}$.  The additional factors $4$ and $2$, that enter the hard
scales, originate from the SL function $\cF_T$ in \eqref{eq:cFTone}
and have a simple kinematical origin.  The standard definition of the
event plane implies that when the secondary gluon is emitted in the
{\em right}\/ hemisphere (see Fig.~\ref{fig:plane}b), all three hard
partons experience equal recoils,
\begin{equation}
\label{eq:I1}
\left.
\begin{split}
k_z>0, \quad & k_y>0: \quad   p_{1x}=-k_x=p_{2x}=-p_{3x}\quad\\
       & k_y<0: \quad   p_{1x}=-k_x=-p_{2x}=p_{3x}\quad
\end{split}
\right\}\> \Longrightarrow \> |k_x|=\frac{\Ko}{4}\,.
\end{equation}
On the other hand, for the secondary gluon in the {\em left}\/
hemisphere (see Fig.~\ref{fig:plane}c and \ref{fig:plane}d), only one
parton recoils against it:
\begin{equation}
\label{eq:I23}
\left. 
\begin{split}
k_z<0, \quad &  k_y>0: \quad   p_{2x}=-k_x; \>\>  p_{1x}=p_{3x}=0\quad \\
             &  k_y<0: \quad   p_{3x}=-k_x; \>\> p_{1x}=p_{2x}=0\quad
\end{split}
\right\}\> \Longrightarrow \> |k_x|=\frac{\Ko}{2}\,.
\end{equation}
These are the factors $4$ or $2$ entering the scales in
\eqref{eq:T-dSigma}.

\EPSFIGURE[ht]{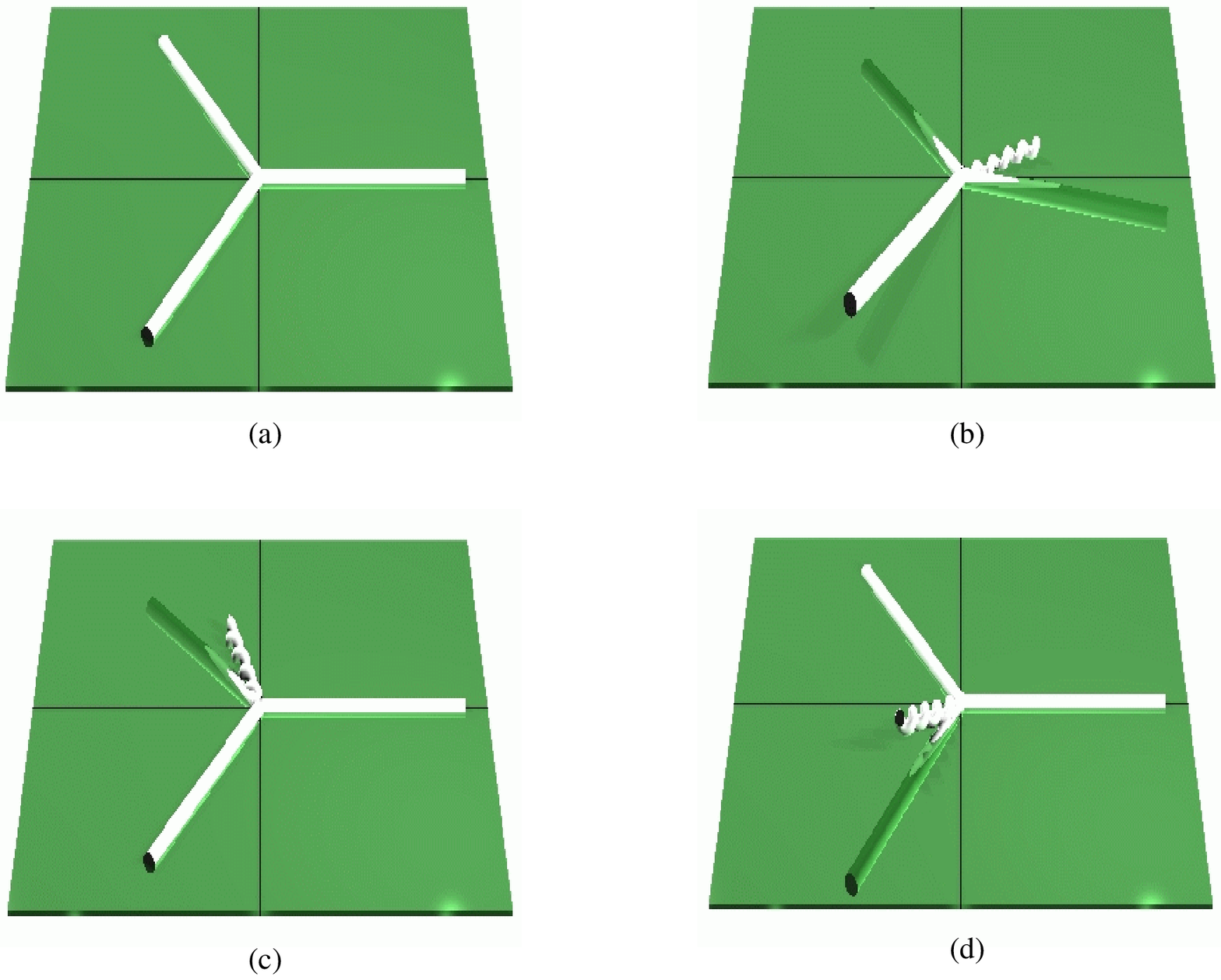,width=0.8\textwidth}{(a) Hard partons in a
  generic Born configuration.
(b) Soft gluon $k$ (the short curly stick) is emitted in the right
hemisphere. All three hard partons experience equal out-of-plane
recoils, see \eqref{eq:I1}.  White (shadowed) partons have positive
(negative) $x$-components of the momentum.
(c) Soft gluon $k$ is emitted in the up-left region. According to
\eqref{eq:I23}, only parton \#2 recoils (shadowed; it has a negative
$x$-components of the momentum).  (d) The case of $k$ in the down-left
region.
\label{fig:plane}}
 
\subsection{$\Ko^R$ case: NP radiator and recollection of PT results}
Here we describe how the results of the previous subsection change in
the case of the $\Ko^R$ distribution.  (In this subsection the
suppressed indices $R$ and $\conf$ are implied.)

\subsubsection{$\Ko^R$-radiator}
\label{sec:rad-R}
As in the previous case, the radiator for the $\Ko^R$ distribution has
PT and NP components:
\begin{equation}
  \label{eq:R-Rad}
  \cR(\nu,\be,q_{1x})=
\cR^{\PT}(\nu,\be)+\de\cR(q_{1x})\>.
\end{equation}
The PT radiator does not depend on $q_{1x}$ while the NP correction
does not depend on $\be$.

The NP correction is described in details in Appendix \ref{App:RadNP}.
Since the observable accumulates only particles emitted in the right
hemisphere which contains only one hard parton $\#1$, the NP radiator
can be written as
\begin{equation}
  \label{eq:R-RadNP}
  \delta \cR(q_{1x})\>=\> 
\nu\>\cp\> C_1 \ln\frac{Q^{\NP}_{1}}{|q_{1x}|}\,.
\end{equation}
Subleading contributions due to radiation into the $R$ hemisphere off
the $L$-hemisphere partons are embodied in the hard NP scale
$Q^{\NP}_{1}\sim{Q}$.  This new scale {\it is not}\/ related to the NP
hard scale \eqref{eq:QNP} for the $\Ko^T$ case.\footnote{Denoting, for
  the sake of simplicity, the hard scales for $R$- and
  $T$-distributions by the same symbol does not cause confusion.}
Contrary to the $\Ko^T$--case, the hard scale in \eqref{eq:cFTone}
does not have a simple geometrical interpretation.
This is due to the fact that particle selection in the $R$-hemisphere
is ``unnatural'' for the two-parton antenna dipoles that determine the
structure of the accompanying soft radiation in the three-jet event.
The procedure for computing the $R$-scale in \eqref{eq:R-RadNP} is
described in Appendix~\ref{App:NPRadR}.

To evaluate the effect of the NP radiator upon the distribution we
need to know the PT radiator. It has been computed in \cite{acopt}
(see also Appendix \ref{App:RadPT}) and reads
\begin{equation}
\label{eq:R-RadPT}
  \cR^{\PT}(\nu\be)=R_1(\rho)=C_1\,r(\rho,Q_1^{\PT})\>,\qquad 
\rho=\bnu\sqrt{1+\be^2}\>,
\end{equation}
with $r$ the DL function defined in \eqref{eq:r}. The hard PT scale
$Q_1^{\PT}\sim Q$ associated with the soft parton emission in the
right region, does {\it not}\/ coincide with the hard scale
\eqref{eq:QPT} for the PT $\Ko^T$-radiator and, once again, does not
have a simple geometrical meaning.

\subsubsection{Recollection of the PT distribution in $\Ko^R$ 
\label{sec:recollationR}}
The situation here is simpler than before since we have a single
Fourier variable $\be$.  Neglecting the NP radiator, the
$q_{1x}$-integral can be freely performed and one obtains
\begin{equation}
  \label{eq:R-SigPT1}
  \Sigma^{\PT}(\Ko)=\int\frac{d\nu\,e^{\nu\Ko}}{2i\pi\nu}\>
\sigma^{\PT}(\nu)\,,
\quad
  \sigma^{PT}(\nu)= \int_{-\infty}^{\infty} 
\frac{d\be}{\pi(1\!+\!\be^2)}\,
e^{-R_1\left(\bnu\sqrt{1+\be^2}\right)}\>.
\end{equation}
By using the identity
\begin{equation}
  \label{eq:R-SigPT2}
  \Sigma^{\PT}(\Ko)=e^{-R_1\left(\bKo^{-1}e^{-\partial_z}\right)}
\frac{f(z)}{\Gamma(1+z)}\>, \qquad 
f(z)=\frac{\Gamma\left(\frac{1+z}{2}\right)}{\sqrt\pi\Gamma(1+\half  z)}\>,
\end{equation}
we obtain the SL expression of the $\Ko^R$ distribution
\begin{equation}
  \label{eq:R-SigPT}
\begin{split}
  & \Sigma^{\PT} (\Ko)\>\simeq \>e^{-R_1\left(\bKo^{-1}\right)}\cdot
  \frac{f(R'_1)}{\Gamma(1+R'_1)}, \qquad R'_1\equiv C_1
  r'\left(\Ko^{-1},Q\right)\>,
\end{split}
\end{equation}
with corrections beyond SL accuracy.
In general we introduce the jet $\#a$ function
\begin{equation}
\label{eq:cFa}
\cF_a(\nu)=e^{R_a(\bnu)} \int_{-\infty}^{\infty}
\frac{d\be_a\,e^{-R_a\left(\bnu\sqrt{1+\be^2_a}\right)}}{\pi(1+\be_a^2)}
\simeq f(R'_a(\nu))\>.
\quad R'_a(\nu)\equiv C_a\,r'(\nu,Q)\,,
\end{equation}
As before, the $\cO{\as}$ expansion of the differential distribution
shows that the SL corrections have a simple kinematical explanation.
From $f(R'_a)\simeq 1-\ln2\cdot R'_a$ we obtain
\begin{equation}
  \label{eq:R-dSigma}
  \frac{d\Sigma^{\PT}(\Ko)}{d\ln \Ko}=\frac{2\as}{\pi}\,
C_{1}\ln\frac{2Q_{1}^{\PT}}{\Ko}+\> \cO{\as^2}\,.
\end{equation}
The additional factor 2 in the scale has the following interpretation.
When a single gluon $k$ is emitted in the $R$-hemisphere (see
Fig.~\ref{fig:plane}b) all four partons, as we already know,
experience equal recoils, $|k_x|$. Counting only the $R$-hemisphere
particles we have $\Ko=2|k_x|$.

\section{NP correction to the $\Ko^T$ distribution and mean \label{sec:T-NP}}
(In this section the suppressed indices $T$ and $\conf$ are implicit,
unless explicitly needed.)
Taking into account the NP radiator \eqref{eq:T-RadNP} we obtain, in
the linear approximation in $1/Q$, (see Appendix~\ref{App:RadNP})
\begin{equation}
  \label{eq:T-measureNP}
\begin{split} 
\sigma(\nu)
&=\!\int d\mu^T\,e^{-\cR(\nu,\be,\gam,q)}
 =\!\int d\mu^T\,e^{-\cR^{\PT}(\nu,\be,\gam)}
\prod_a\left(\frac{|q_{ax}|}{Q_{a}^{\NP}}\right)^{\nu\, C_a\cp}\\
&=\!\int d\mu^T e^{-\cR^{\PT}(\nu,\be,\gam)}
\left\{1\!-\!\nu\cp \sum_a C_{a}
\left[\,\ln(\bnu Q_{a}^{\NP})\!+\!\tchi_a(\be,\gam)\,\right]
\right\}.
\end{split}
\end{equation}
The unity in the curly brackets gives the PT contribution, while the
second term describes the leading NP correction.  The functions
$\tchi_a$ are given in Appendix~\ref{App:RadNP}.  They depend only on
the Fourier variables and are constructed with use of the function
$\chi(\be)$:
\begin{equation}
  \label{eq:chi}
\chi(\be)\!=\!\ln\sqrt{1\!+\!\be^2}\!+\!\be\tan^{-1} \be=
\abs{\be}\frac{\pi}{2}\!+\!\chi'(\be)\,,
\qquad \int_{-\infty}^\infty\frac{d\be}{\pi(1\!+\!\be^2)}\,\chi'(\be)= 0\,,
\end{equation}
with $\chi'(\be)=\cO{\ln|\be|}$ for large $\be$. For $a\!=\!2,3$ we
have $\tchi_a=\chi(\be_a)$. The function $\tchi_1(\be,\gam)$ contains
the combination of $\chi(\be_{12})$ and $\chi(\be_{13})$ and is
defined in \eqref{eq:tchi}.

Equation~\eqref{eq:chi} shows that the functions $\tchi_a$ linearly
grow for large $\abs{\be_a}$, and this makes the structure of the NP
corrections quite different from that of the PT contributions.
Indeed, for the PT contribution the Fourier variable integrations in
$d\mu^T$ are rapidly convergent, so that all soft momenta are of order
$\Ko$.
For the NP corrections, due to the presence of the $\tchi_a$
functions, the Fourier variable integrations in $d\mu^T$ become
logarithmic. This implies that the phase space regions with the
secondary gluon momenta much smaller than $\Ko$ become important.

From \eqref{eq:T-sigPT} and \eqref{eq:T-measureNP} we obtain
\begin{equation}
  \label{eq:T-sigma}
\sigma(\nu)=
\sigma^{\PT}(\nu)-\nu\,\cp\sum_a\,C_a\>f_a(\nu)\>,
\end{equation}
where
\begin{equation}
  \label{eq:fa}
\begin{split}
&f_{a}(\nu)\!=\! \sigma^{\PT}(\nu)\cdot\ln(\bnu Q_a^{\NP})
+\tf_a(\nu)\,,\quad
\tf_a(\nu)\!=\!\int d\mu^T\>\tchi_a(\be,\gam)\>e^{-\cR^{\PT}(\bnu,\be,\gam)}\>.
\end{split}
\end{equation}
The functions $\tf_a$ are analyzed in Appendix~\ref{App:tfa}. We show
that they are given by an integral similar to the one for the PT
contribution to the Mellin moment, except for the presence of the
$\chi$ functions in the integrand: 
\begin{equation}
  \label{eq:tfa}
\begin{split}  
\sigma^{\PT}(\nu)&=\int_{-\infty}^{\infty}\prod_{a=1}^3 
\frac{d\be_a}{\pi(1+\be_a^2)}\>
\cS(\nu,\be_1,\be_{23})\>e^{-R_2\left(\bnu\sqrt{1+\be^2_2}\right)
-R_3\left(\bnu\sqrt{1+\be^2_3}\right)}\>,\\
 \tilde f_a(\nu)&=\int_{-\infty}^{\infty}\prod_{a=1}^3 
\frac{d\be_a}{\pi(1+\be_a^2)}\>
\cS(\nu,\be_1,\be_{23})\>e^{-R_2\left(\bnu\sqrt{1+\be^2_2}\right)
-R_3\left(\bnu\sqrt{1+\be^2_3}\right)}
\cdot \chi(\be_a)\>,
\end{split}
\end{equation}
where $\be_{23}\!=\!\be_2\!+\!\be_3$.  The factor $\cS$ is the
contribution of the parton $\#1$ (see Appendix~\ref{App:RadPT}) and is
given by
\begin{equation}
  \label{eq:cS}
\cS (\nu,\be,\be_{23})\>=\>e^{-\bR_1(\bnu,\be,\bbe,0)} -\frac{2}{\pi} 
\int_{0}^\infty\frac{d\gam}{\gam} \>
\Im\, e^{-\bR_1(\bnu,\be,\bbe,\gam)}\>, \quad \bbe\equiv \be\!+\!\be_{23}\>,
\end{equation}
with $\bR_1$ defined in \eqref{eq:T-RadPT}. 

The presence of the function $\chi(\be_a)$, linearly growing with
$|\be_a|$, in the integrand of \eqref{eq:tfa}, makes the evaluation of
$\tf_a(\nu)$ quite involved.  The large values of $\be_a,\gam$,
corresponding to the recoil momenta much smaller than $\Ko$, give the
dominant contribution to $\tf_a$.
In these circumstances we cannot use the SL expansion of the PT
radiator \eqref{eq:SL-Rad} which simplified the calculation for finite
values of the Fourier variables.

The strategy of the calculation is the following. We split
$\chi(\be_a)$ into two pieces \eqref{eq:chi}.  The contribution of
$\chi'(\be_a)$ is evaluated by the same techniques that we used for
the PT case. Indeed, since $\chi'(\be_a)$ is growing only
logarithmically, the Fourier variable integrations remain convergent
and we can use the SL expansion \eqref{eq:SL-Rad}.
The linear piece $\frac{\pi}{2}|\be_a|$ which gives the leading
contribution to $\tf_a(\nu)$ will be analyzed separately.

\paragraph{Result for $\tf_1$.} 
In this case the $\be_1$--integration is logarithmic while the
remaining $\gam, \be_2,\be_3$ integrals remain convergent. To analyze
the leading piece we need to study the behaviour of $\cS$ for large
$\be_1$ and finite $\gam, \be_2,\be_3$.  In this limit the imaginary
part of $\bR_1$ vanishes, and $\cS$ reduces to the first term only,
see Appendix \ref{App:tf1},
\begin{equation}
  \label{eq:cS1}
\cS(\nu,\be_1,\bbe)\simeq 
e^{-R_1(\rho_1)}\>,\qquad \rho_{12},\,\rho_{13}\simeq
\rho_1\equiv\bnu\sqrt{1+\be_1^2}\>.
\end{equation}
This behaviour can be simply understood by considering the event-plane
kinematics \eqref{eq:I1}, \eqref{eq:I23}. For $\be_1\to\infty$ we are
forcing parton $\#1$ to stay in the plane by suppressing PT radiation
off it.  At the same time, radiation off the other two partons is not
constrained, since parton $\#1$ does not recoil when PT emission
occurs in the $L$-hemisphere.  As a result, $\cS$ turns out to be
simply the Sudakov suppression factor for not having PT gluon
radiation off the parton $\#1$.

The leading piece of $\tf_1$ is obtained by substituting
$\chi(\be_1)\to \frac{\pi}{2}|\be_1|$ in \eqref{eq:tfa} and using the
approximation \eqref{eq:cS1}. All $\be_a$ integrals are now factorized
and we obtain
\begin{equation}
  \label{eq:tf1as}
\begin{split}
  \tf_1(\nu)&\simeq\prod_{a=2,3}\int_{-\infty}^{\infty}
  \frac{d\be_a\,e^{-R_a\left(\bnu\sqrt{1+\be^2_a}\right)}}{\pi(1+\be_a^2)}
  \int_{\bnu}^{\infty}\frac{d\rho_1}{\rho_1}\,e^{-R_1(\rho_1)}\\
  &= \prod_{a=1}^3 e^{-R_a(\bnu)}\>\cF_2(\nu)\cF_3(\nu)\>E_1(\bnu)\>,
\end{split}
\end{equation}
with $\cF_a(\nu)$ the SL functions defined in \eqref{eq:cFa}.  The
function $E_1(\bnu)$ given by
\begin{equation}
  \label{eq:E1}
  E_1(\bnu)=\int_{\bnu}^{\infty}\frac{d\rho}{\rho}\>
e^{-R_1(\rho)+R_1(\bnu)}
\end{equation}
has been studied in \cite{broad}. Its properties are recalled in
Appendix~\ref{App:E}.

Corrections to the leading behaviour of $\tf_1$ are discussed in
Appendix~\ref{App:tfa}. The SL result reads
\begin{equation}
  \label{eq:tf1}
  \tf_1(\nu)= e^{-\sum_a R_a(\bnu)}
\left\{\,\cF_2(\nu)\cF_3(\nu)\>E_1(\bnu)+\cC_1(\nu)\,\right\}\>,
\end{equation}
with $\cC_1(\nu)$ a SL function which vanishes for $\nu=0$ and is
proportional to $r'(\nu,Q)$.

The function $E_1(\Ko^{-1})$ is related to the {\em restricted}\/
average of $\ln{Q_1^{\NP}}/{|q_{1x}|}$, under the condition
$|q_{1x}|\le \Ko$.  To compute this average, we use the Sudakov
distribution in the recoil momentum $q=|q_{1x}|$ of parton $\#1$,
\begin{equation}
  \label{eq:D1}
  \cD_1(q)=\frac{d} {d\ln q}\,e^{-R_1(q^{-1})}\,,
\end{equation}
and derive
\begin{equation}
  \label{eq:L1}
\begin{split}
L_1(\Ko)&\equiv
\VEV{\ln\frac{Q_1^{\NP}}{|q_{1x}|}}_{|q_{1x}|\le\Ko}\!\!
\!=\!e^{R_1(\Ko^{-1})}\!\int_0^{\Ko}\frac{dq}{q}\,\cD_1(q)\>
\ln\frac{Q_1^{\NP}}{q}
\\ &
= \ln\frac{Q_1^{\NP}}{\Ko}+E_1(\Ko^{-1})\>.
\end{split}
\end{equation}
From the behaviour of the function $E_1$, see \eqref{eq:regione1} and
\eqref{eq:regione2}, we obtain the leading terms of $L_1$
\begin{equation}
  \label{eq:L1as}
\begin{split} 
& L_1(\Ko)=\ln\frac{Q_1^{\NP}}{\Ko}
\left(1+\cO{\left(\as\ln^2 \frac{Q}{\Ko}\right)^{-1}}\right),
\qquad \as\ln^2{Q}/{\Ko}\gg1\>,\\
& L_1(\Ko)=\VEV{\ln\frac{Q_1^{\NP}}{|q_{1x}|}}
\left(1+\cO{\as\ln^2\frac{Q}{\Ko}}\right), \qquad \as\ln^2Q/\Ko\ll1\>.
\end{split}
\end{equation}
The {\em unrestricted}\/ average is given by
\begin{equation}
\label{eq:VEV1}
\VEV{\ln\frac{Q_1^{\NP}}{|q_{1x}|}}=
E_1\left(\frac{1}{Q_{1}^{\NP}}\right)=
\frac{\pi}{2\sqrt{C_1\as(Q)}}-\ln \frac{Q_1^{\PT}}{Q_1^{\NP}}
 -\frac{\be_0}{6C_1} +\cO{\sqrt{\as}}\>.
\end{equation}
Up to corrections of order $\sqrt{\as}$ it is independent of the event
geometry.

The restricted average $L_1(\Ko)$ is a decreasing function of
$\Ko$. It starts from $\ln Q_1^{\NP}/\Ko$ for very small $\Ko$
($\as\ln^2{Q}/{\Ko}\gg1$). With $\Ko$ increasing, it freezes at the
value of the unrestricted average $E_1({1}/{Q_{1}^{\NP}})$ at
moderately small $\Ko$ ($\as\ln^2{Q}/{\Ko}\ll1$), up to corrections of 
$\cO{\sqrt{\as}}$. 

\paragraph{Result for $\tf_a$ with $a=2,3$.} 
Consider for example the case of $\tf_2$. Here the $\be_2$ integration
is logarithmic while the $\be_1,\be_3$ integrals converge.  Evaluation
of the leading behaviour of $\cS$ for $\be_1,\be_3 \ll \be_2$ is
subtle.
In this $\be$--region the second term of $\cS$ in \eqref{eq:cS} is not
negligible any more since the $\gam$--integration turns out to be
logarithmic.  As shown in Appendix~\ref{App:tfa}, the leading
exponential contributions $\exp(-\half R_1)$ in the two terms of $\cS$
cancel, and one is left with
\begin{equation}
  \label{eq:cS2}
  \cS(\nu,\be,\be_{23})\simeq
e^{-R_1\left(\bnu\sqrt{1+\be_2^2}\right)} .
\end{equation}
This behaviour has the following physical explanation.  For
$\be_2\to\infty$ we force $p_2$ to stay in the plane. From
\eqref{eq:I1} we see that in this case also $p_1$ has to be devoid of
PT radiation, since gluon emission off parton \#1 would cause parton
\#2 recoiling.  Therefore, the distribution for large $\be_2$ is given
by the product of the {\em two}\/ Sudakov form factors: one from
parton $\#2$, giving $\exp(-R_2(\bnu\sqrt{1+\be_2^2}))$ and the other
one from the parton $\#1$ giving \eqref{eq:cS2}.  Radiation off the
parton $\#3$ is not constrained.

The leading piece of $\tf_2$ is obtained by substituting in
\eqref{eq:tfa} $\chi(\be_2)\to \frac{\pi}{2}|\be_2|$ and using the
approximation \eqref{eq:cS2}. All $\be_a$ integrals in \eqref{eq:tfa}
are now factorized. The $\be_1$--integral gives $1$; the remaining two
integrals produce
\begin{equation}
  \label{eq:tf2as}
\begin{split}
  \tf_2(\nu)&\simeq\int_{-\infty}^{\infty}
  \frac{d\be_3\,e^{-R_3\left(\bnu\sqrt{1+\be^2_3}\right)}}{\pi(1+\be_3^2)}
  \int_{\bnu}^{\infty}\frac{d\rho}{\rho}\,e^{-R_1(\rho)}e^{-R_2(\rho)}\\
  &= \prod_{a=1}^3 e^{-R_a(\bnu)}\>\cF_3(\nu)\>E_2(\bnu)\>,
\end{split}
\end{equation}
where we used $\rho=\bnu\sqrt{1+\be_2^2}$.  The function $E_2$ is
defined as 
\begin{equation}
  \label{eq:E2}
  E_a(\bnu)=\int_{\bnu}^{\infty}\frac{d\rho}{\rho}\>
e^{-R_1(\rho)-R_a(\rho)}e^{R_1(\bnu)+R_a(\bnu)}\>
\end{equation}
with $a=2$ and is analyzed in Appendix~\ref{App:Ea}.  
Corrections to the leading behaviour of $\tf_2$ are discussed in
Appendix~\ref{App:tfa}. The answer to SL accuracy reads
\begin{equation}
  \label{eq:tf2}
  \tf_2(\nu)= e^{-\sum R_a(\bnu)}
\left\{\,\cF_3(\nu)\>E_2(\bnu)+\cC_2(\nu)\,\right\}\>,
\end{equation}
with $\cC_2(\nu)$ a SL function which vanishes for $\nu=0$ and is
proportional to $r'(\nu,Q)$.  The expression for $\tf_3$ is obtained
from \eqref{eq:tf2} by interchanging the labels \#2 and \#3.

The functions $E_a(\Ko^{-1})$ with $a=2,3$ can be related, as above,
with the average of $\ln({Q_a^{\NP}}/{|q_{ax}|})$ under the condition
$|q_{ax}|\le \Ko$:
\begin{equation}
  \label{eq:L23}
\begin{split}
L_a(\Ko)&\equiv \VEV{\ln\frac{Q_a^{\NP}}{|q_{ax}|}}_{|q_{ax}|\le\Ko}
=e^{R_1(\Ko^{-1})+R_a(\Ko^{-1})}
\int_0^{\Ko}\frac{dq}{q}\,\cD_a(q)\>\ln\frac{Q_a^{\NP}}{q} \\
&=\ln\frac{Q_a^{\NP}}{\Ko}+E_a(\Ko^{-1})\>,
\end{split}
\end{equation}
where the PT distribution over the parton recoil $|q_{ax}|$ is given
by (the logarithmic derivative of) the product of the Sudakov form
factors of partons $\#1$ and $\#a$:
\begin{equation}
  \label{eq:D23}
  \cD_a(q)=\frac{d} {d\ln q}\,e^{-R_1(q^{-1})-R_a(q^{-1})}\,.
\end{equation}
From the behaviour of the functions $E_a$
(see \eqref{eq:regione1a} and \eqref{eq:regione2a}) we have
\begin{equation}
  \label{eq:L23as}
\begin{split} 
  & L_a(\Ko)=\ln\frac{Q_a^{\NP}}{\Ko} \left(1+\cO{\left(\as\ln^2
        \frac{Q}{\Ko}\right)^{-1}}\right)\>,
  \qquad \as\ln^2{Q}/{\Ko}\gg1\>,\\
  & L_a(\Ko)=\VEV{\ln\frac{Q_a^{\NP}}{|q_{ax}|}}
  \left(1+\cO{\as\ln^2\frac{Q}{\Ko}}\right), \qquad
  \as\ln^2Q/\Ko\ll1\>,
\end{split}
\end{equation}
where the unrestricted averages now read
\begin{equation}
\label{eq:VEV23}
\begin{split}
  \VEV{\ln\frac{Q_a^{\NP}}{|q_{ax}|}}
  &\!=\!E_a\left(\!\frac{1}{Q_{a}^{\NP}}\!\right) \\& \!\simeq\!
  \frac{\pi}{2\sqrt{(C_1\!+\!C_a)\as(Q)}}
  \!-\!\frac{1}{C_1+C_a}\left\{\!C_1\ln\frac{Q_1^{\PT}}{Q_a^{\NP}}
    \!+\!C_a\ln \frac{Q_a^{\PT}}{Q_a^{\NP}}
\!+\!\frac{\be_0}{6}\!\right\},
\end{split}
\end{equation}
where we dropped corrections of absolute order $\sqrt{\as}$ (relative
corrections $\cO{\as}$).  The diagonal ratios $Q_a^{\PT}/Q_a^{\NP}$
include only hard corrections to the scales. The non-diagonal ratios
$Q_1^{\PT}/Q_a^{\NP}$, on the contrary, depend also on the event
geometry.

\subsection{Correction to the mean $\Ko^T$}
The NP correction to the mean is obtained from \eqref{eq:mean} and
\eqref{eq:T-sigma} by taking formally the limit $\nu\to0$ of the
functions $f_a(\nu)$, which actually means $\nu Q=\cO{1}$. We get
\begin{equation}
  \label{eq:T-mean1}
\VEV{\Ko^R}^{\NP}=\cp\,\sum_a C_a\>f_a(0)\>,
\qquad
f_a(0)=\lim_{\nu\to0}\left\{\ln(\bnu Q_a^{\NP})+E_a(\bnu)\right\},
\end{equation}
where we used the fact that, for $\nu=\cO{1/Q}$, we have
$R_a(\bnu)\to0$, $\cF_a(\nu)\to1$, $\cC_a(\nu)\to0$ and
$\sigma^{\PT}(\nu)\to1$.  The quantity in the curly brackets is
$L_a(\bnu^{-1})$.  We obtain, up to corrections of order $\sqrt{\as}$,
\begin{equation}
  f_a(0)\simeq \VEV{\ln\frac{Q_a^{\NP}}{|q_{ax}|}},
\end{equation}
with the average given in \eqref{eq:VEV1} and \eqref{eq:VEV23}.
The final result for the mean is then
\begin{equation}
  \label{eq:T-mean2}
\VEV{\Ko^T}^{\NP} =\cp\sum_a\,C_a\,
\VEV{\ln\frac{Q_a^{\NP}}{|q_{ax}|}}
+\cO{\sqrt{\as}}\>.
\end{equation}
By using the expansions \eqref{eq:VEV1} and \eqref{eq:VEV23} we get
\begin{equation}
  \label{eq:T-mean}
  \VEV{\Ko^T}^{\NP} =\cp\frac{\pi}{2\sqrt{\as(Q)}}
\left(\sqrt{C_1}+\frac{C_2}{\sqrt{C_1+C_2}}+\frac{C_3}{\sqrt{C_1+C_3}}\right)
+\cO{1}\>.
\end{equation}
The legitimate constant term here can be restored from \eqref{eq:VEV1}
and \eqref{eq:VEV23}; it contains the ratios of hard PT and NP scales
and depends on the geometry of the event.

The peculiar colour structure of \eqref{eq:T-mean} shows that, due to
an interplay between the PT and NP effects, the three jets do not
contribute independently to the mean.

\subsection{The distribution \label{subsec:distribution}}
From the Mellin moment \eqref{eq:T-sigma} one obtains the NP
correction to the integrated distribution
\begin{equation}
 \Sigma(\Ko)= \Sigma^{\PT}(\Ko)+\de \Sigma(\Ko)
\end{equation}
where
\begin{equation}
  \!\!\de \Sigma(\Ko)=-\cp\sum_aC_a \cdot \psi_a(\Ko)\>,
\quad
\psi_a(\Ko)=\partial_{\Ko}\!\!\int\frac{d\nu}{2\pi i \nu}e^{\nu\Ko}f_a(\nu),
\end{equation}
with $f_a(\nu)$ given in \eqref{eq:fa}. 
To evaluate the Mellin integral we use the operator identity
\begin{equation}
  \label{eq:psi-a}
\begin{split}  
&\psi_a(\Ko) 
\!=\!\int d\mu^T e^{-\cR^{\PT}(e^{-\partial_z},\be,\gam)}\>
\left\{\ln Q_a^{\NP}-\partial_z+\tchi_a(\be,\gam)\right\}
\left. \frac{\partial_{\Ko}\bKo^{z}}{\Gamma(1+z)}\right|_{z=0}\\
&\!\!\!\!=\!\!\int\! d\mu^T e^{-\cR^{\PT}(e^{-\partial_z},\be,\gam)}\!
\left\{\!\ln
  \frac{Q_a^{\NP}}{\bKo}+\psi(1+z)-\frac{1}{z}+\tchi_a(\be,\gam)\!\right\}
\left. \frac{\partial_{\Ko}\bKo^{z}}{\Gamma(1+z)}\right|_{z=0}, 
\end{split}
\end{equation}
where in the first line we have replaced $\ln \nu$ with $-\partial_z$.

Consider the two pieces of $\psi_a$ given by the first two and the
last two terms in the curly brackets on the second line of
\eqref{eq:psi-a}.  The first piece can be expressed in terms of the PT
distribution as follows:
\begin{equation}
  \label{eq:psi-1}
\begin{split}
\psi^{(1)}_a(\Ko)\simeq 
\Delta_a^{(1)}\cdot \partial_{\Ko} \Sigma^{\PT}(\Ko)\>,
\end{split}
\end{equation}
where
\begin{equation}
    \label{eq:T-Delta1}
\Delta_a^{(1)}=\ln\frac{Q_a^{\NP}}{\bKo}+\psi(1+R'_T)=
\ln\frac{Q_a^{\NP}}{\Ko}\>\left(1+\cO{\as}\right).
\end{equation}
Here we used the operator identity \eqref{eq:identity} in the form
\begin{equation}
  \label{eq:identity1}
\Sigma^{\PT}(\Ko)=\int d\mu^T e^{-\cR^{\PT}(e^{-\partial_z},\be,\gam)}\>
\left. \frac{\bKo^{z}}{\Gamma(1+z)}\right|_{z=0}\>,
\end{equation}
and substituted $R_T'$ for $z$ with corrections of order $\cO{\as}$.
The contribution \eqref{eq:psi-1} to the NP correction
$\de\Sigma(\Ko)$ clearly has the form of a shift of the PT
distribution.

In a similar way we calculate the second contribution to $\psi_a(\Ko)$
given by the last two terms in the curly brackets of \eqref{eq:psi-a}.
Neglecting corrections of order $\as$ we have\footnote{The
  substitution of $z$ with $R_T'$ is in general correct only for
  regular functions of $z$. Actually here we can apply the
  substitution for the function $1/z$ since the singularity cancels.
  With more cumbersome algebra one can avoid this intermediate step
  altogether.}
\begin{equation}
  \label{eq:psi-2}
\begin{split}
\psi^{(2)}_a(\Ko)
\simeq -\frac{1}{R_T'}\>\partial_{\Ko}\Sigma^{\PT}(\Ko) \>+\>
\frac{\partial_{\Ko}\tf_a\left(\Ko^{-1}\right)}{\Gamma\left(1+R'_T\right)}.
\end{split}
\end{equation}
Let us show that the second term with the derivative of $\tf_a$
generates a $1/R_T'$ contribution which cancels the singularity at
$R_T'=0$ of the first term in \eqref{eq:psi-2}.
To verify this cancellation we consider the singular pieces of $\tf_a$
containing the $E_a$ functions (see \eqref{eq:tf1as} and
\eqref{eq:tf2as}).

We start from the leading piece of $\tf_1$ in \eqref{eq:tf1as} and
take the derivative to obtain
\begin{equation}
  \label{eq:tf1derivative'}
\!\!\!\partial_{\Ko}\tf_1\left(\Ko^{-1}\right)
\simeq \frac{e^{-\sum R_a\left(\bKo^{-1}\right)}}{\Ko}\!
\cF_2\left(\Ko^{-1}\right)\!\cF_3\left(\Ko^{-1}\right)
\left\{1\!+\!(R'_2\!+\! R'_3)E_1\left(\bKo^{-1}\right)\>\right\}.
\end{equation}
Using \eqref{eq:T-SigPT}
and \eqref{eq:cFa}, we
arrive at
\begin{equation}
  \label{eq:tf1derivative}
 \frac{\partial_{\Ko}\, \tf_1\left(\Ko^{-1}\right)}{\Gamma(1+R'_T)}
\>\simeq\> \frac{\cF_2\cF_3} {\cF_T}
\left\{\frac{C_2+C_3}{C_T}\,E_1\>+\>\frac{1}{R'_T}\right\}
\cdot\left(\partial_{\Ko}\Sigma^{\PT}\right),
\end{equation}
with all functions in the r.h.s.\ evaluated at $\bKo^{-1}$. The second
term is singular for $R'_T\to0$ and cancels with the $-1/R_T'$ term in
\eqref{eq:psi-2}. (Notice that $\cF_T$ and $\cF_a$ tend to unity in
this limit.)

Similar results are obtained from the leading pieces of $\tf_2$ and
$\tf_3$. Taking the derivative of \eqref{eq:tf2as} we obtain
\begin{equation}
  \label{eq:tf2derivative}
\frac{\partial_{\Ko}\,\tf_2\left(\Ko^{-1}\right)}{\Gamma(1+R'_T)}
\>\simeq\>\frac{\cF_3}{\cF_T} \left\{\frac{C_3}{C_T}\,E_{2}
\>+\>\frac{1}{R'_T}\right\}
\cdot\left(\partial_{\Ko}\Sigma^{\PT}\right),
\end{equation}
and again the $-1/R'_T$  singularity in \eqref{eq:psi-2} cancels.

The derivative of the subleading pieces $\cC_a$ in the expressions
\eqref{eq:tf1-fine} and \eqref{eq:tf2-fine} for $\tf_a$ is simpler to
evaluate. We have
\begin{equation}
  \label{eq:DeltaC}
 \partial_{\Ko}
\left(\frac{\cC_a\,e^{-\sum_b R_b}}{\Gamma(1+R'_T)}\right)
\>\simeq\> 
\frac{\cC_a}{\cF_T}
\partial_{\Ko}\Sigma^{\PT}(\Ko)\>,
\end{equation}
where all functions are evaluated at $\bKo^{-1}$ and we used again
\eqref{eq:T-SigPT}.

In conclusion, the second piece $\psi_a^{(2)}(\Ko)$ of the NP
corrections of the $\Ko^T$-distribution can also be written as a shift
of the PT distribution. Up to corrections of relative order $\as$, we
have
\begin{equation}
  \label{eq:psi-22}
  \psi_a^{(2)}(\Ko)\simeq \Delta_a^{(2)}\cdot 
\partial_{\Ko} \Sigma^{\PT}(\Ko)\>,
\end{equation}
where 
\begin{equation}
  \label{eq:Delta-2}
\begin{split}
  &\Delta_1^{(2)}\>\simeq\>
  \frac{C_2+C_3}{C_T}\frac{\cF_2\cF_3}{\cF_T}\,E_1
  \>+\>\frac{\cF_2\cF_3-\cF_T}{\cF_T\,R'_T}\>+\>\frac{\cC_1}{\cF_T}\>,\\
  &\Delta_2^{(2)}\>\simeq\> \frac{C_3\cF_3}{C_T\cF_T}\,E_{2}
  \>+\>\frac{\cF_3-\cF_T}{\cF_T\,R'_T}\>+\>\frac{\cC_2}{\cF_T}\>,\\
  &\Delta_3^{(2)}\>\simeq\> \frac{C_2\cF_2}{C_T\cF_T}\,E_{3}
  \>+\>\frac{\cF_2-\cF_T}{\cF_T\,R'_T}\>+\>\frac{\cC_3}{\cF_T}\>,
\end{split}
\end{equation}
with all functions evaluated at $\bKo^{-1}$.
The first terms are the leading pieces. 
The second terms are SL functions regular for $R'_T\to0$ (the
numerators vanish when $r'\to0$).
The last terms are SL corrections which vanish with $r'\to0$.

Summing up the two corrections \eqref{eq:psi-1} and \eqref{eq:psi-22},
$\Delta_a=\Delta_a^{(1)}+\Delta_a^{(2)}$, we finally obtain the
expression for the NP shift
\begin{equation}
  \label{eq:T-shift}
\begin{split}
\Sigma(\Ko)\>=\>
\Sigma^{\PT}(\Ko)+\delta\Ko\cdot \partial_{\Ko}\Sigma^{\PT}(\Ko)
\>\simeq \>\Sigma^{\PT}(\Ko\!+\!\delta\Ko)\>,
\end{split}
\end{equation}
with 
\begin{equation}
  \label{eq:dKoT}
\delta\Ko=-\cp\sum_a C_a\,\Delta_a(\Ko)\>.
\end{equation}
By expanding $\Delta^{(1)}_a$ and $\Delta^{(2)}_a$ and using the
functions $L_a(\Ko)$ given in \eqref{eq:L1}, \eqref{eq:L23} and
describing the restricted averages, we arrive at
\begin{equation}
  \label{eq:Delta}
\begin{split}  
  &\Delta_1(\Ko)\simeq
  \frac{C_2+C_3}{C_T}\,\bar{L}_1(\Ko)\>+\>\frac{C_1}{C_T}\ln\frac{4Q_1^{\NP}}{\Ko}\>,\\
  &\Delta_2(\Ko)\simeq \frac{C_3}{C_T}\,\bar
  L_2(\Ko)\quad\>+\>\frac{C_1}{C_T}\ln\frac{4Q_2^{\NP}}{\Ko}
  \> +\>\frac{C_2}{C_T}\ln\frac{2Q_2^{\NP}}{\Ko}\>,\\
  &\Delta_3(\Ko)\simeq \frac{C_2}{C_T}\,\bar
  L_3(\Ko)\quad\>+\>\frac{C_1}{C_T}\ln\frac{4Q_3^{\NP}}{\Ko}
  \>+\>\frac{C_3}{C_T}\ln\frac{2Q_3^{\NP}}{\Ko}\>.
\end{split}
\end{equation}
Here we introduced the function $\bar{L}_a(\Ko)=L_a(\Ko)-\gamma_E$ and
we neglected corrections of relative order $r'=\cO{\as\ln Q/\Ko}$.

To understand the structure of the result we consider two regimes.

\paragraph{$\as\ln^2Q/\Ko\gg1$.} This regime corresponds to
well-developed PT cascades.  From \eqref{eq:L1as} and \eqref{eq:L23as}
we have $L_a(\Ko)\sim\ln(Q/\Ko)$, so that the colour factors in
\eqref{eq:Delta} cancel and all $\Delta_a(\Ko)\sim\ln(Q/\Ko)$.
For the shift we then simply have
\begin{equation}
  \label{eq:shitfT-well}
\de\Ko \simeq -\cp\,\sum_a C_a\>\left(\ln\frac{Q}{\Ko} + \cO{1}\right).
\end{equation}
In this regime multiple PT radiation causes all three hard partons to
experience comparable recoil $|q_{ax}|\sim \Ko$.  The NP radiation off
each of the partons then averages to $\ln Q_a^{\NP}/|q_{ax}|\simeq \ln
Q_a^{\NP}/\Ko$ thus producing the above result.

\paragraph{$\as\ln^2Q/\Ko\ll1$.} In this regime there are only few
secondary PT partons. Let us restrict ourselves to the first-order in
$\as$, that is to emission of a single PT gluon:
\begin{equation}
  \label{eq:deSigma}
\Sigma-\Sigma^{\PT}\simeq
\de\Ko\>\frac{d\Sigma^{\PT}}{d\Ko}\simeq \de\Ko\>\frac{R'_T}{\Ko}\>
\Sigma^{\PT} 
\>\simeq\> \de\Ko\>\frac{R'_T}{\Ko}
\>.
\end{equation}
Observing that $R_T'\propto C_T$, we can represent the characteristic
product $\de\Ko\>R'_T$ in the following equivalent form:
\begin{equation}
  \label{eq:T-shift1}
  \de\Ko\>R'_T=\sum_{a=1}^3\>R'_a(\Ko^{-1})\>\de\Ko^{(a)}\>,
\end{equation}
where (up to $\as$ corrections) 
\begin{equation}
  \label{eq:deKa}
\begin{split}
  &\de \Ko^{(1)}=-\cp\left\{
    C_1\ln\frac{4Q_1^{\NP}}{\Ko}+C_2\ln\frac{4Q_2^{\NP}}{\Ko}+
    C_3\ln\frac{4Q_3^{\NP}}{\Ko}\right\},\\
  &\de \Ko^{(2)}=-\cp\left\{
    C_1\,\bar L_1(\Ko)+C_2\ln\frac{2Q_2^{\NP}}{\Ko}+C_3\,\bar L_3(\Ko)\right\},\\
  &\de \Ko^{(3)}=-\cp\left\{ C_1\,\bar L_1(\Ko)+C_2\,\bar
    L_2(\Ko)+C_3\ln\frac{2Q_3^{\NP}}{\Ko}\right\}.
\end{split}  
\end{equation}
Being accompanied by the factor $R'_a(\Ko^{-1})\propto C_a$ in
\eqref{eq:T-shift1}, the contribution $\de \Ko^{(a)}$ to the NP shift
corresponds to the PT gluon emission off the parton $\#a$.

Thus, the shift $\de\Ko^{(1)}$ given by the first line in
\eqref{eq:deKa} arises when the PT gluon is emitted from $p_1$. In
this case, as we know from \eqref{eq:I1}, all partons undergo equal
recoils, so that $|q_{ax}|\sim{\Ko}/{4}$.  Each hard parton then
contributes with $C_a\ln{Q_a^{\NP}}/{|q_{ax}|}\simeq
C_a\ln(4Q_a^{\NP})/\Ko$.

The shift $\de\Ko^{(2)}$ arises when the PT gluon is emitted from
$p_2$. In this case only parton $p_2$ recoils with
$|q_{2x}|={\Ko}/{2}$ (see \eqref{eq:I23}).
In this situation, the NP gluon emitted from $p_2$ contributes to the
shift, as before, $C_2\ln{Q_2^{\NP}}/{|q_{2x}|}\simeq
C_2\ln(2Q_2^{\NP})/\Ko$ (cf.\ the second term on the second line in
\eqref{eq:deKa}).  When the NP radiation occurs off \#1 and \#3 the
situation is entirely different.  Since the partons \#1 and \#3 do not
recoil against the PT gluon emission in the region \#2 (up-left
quadrant), the momenta $p_1$ and $p_3$ stay in the event plane. In
these circumstances the NP radiation from \#1 and \#3 would diverge if
not for the high-order Sudakov form factor effects that push the
partons off the plane by certain $q_{x}\neq0$.  The corresponding
averages of $\ln{Q_a^{\NP}}/{|q_{ax}|}$, $a=1,3$ give the singular
$L$-factors \eqref{eq:L1} and \eqref{eq:L23}.  In a similar way one
can interpret various terms in $\de\Ko^{(3)}$.

We notice in conclusion that, using \eqref{eq:T-shift}, the
$\Ko^T$-distribution with account of the leading NP correction can be
written in the form
\begin{equation}
  \label{eq:T-shift'}
  \Sigma(\Ko)\simeq e^{-\sum_a R_a\left(\bar K_a^{-1}\right)}
\frac{\cF_T(\Ko^{-1})}{\Gamma(1+R'_T)}\>,
\quad \bar K_a=e^{-\gam_E}(\Ko+\de\Ko^{(a)})\>.
\end{equation}
where, up to correction of order $r'$, $\de\Ko^{(a)}$ are given in
\eqref{eq:deKa}.

\section{NP correction for the $\Ko^R$ distribution and mean \label{sec:R-NP}}
Here and in the rest of this section the indices $R$ and $\conf$ are
implied.
The integration over the recoil $q_{1x}$ gives (see \eqref{eq:T-INP}
and \eqref{eq:tchi})
\begin{equation}
  \label{eq:R-measureNP}
\begin{split}
  \sigma(\nu) &=\int
  d\mu^R\,e^{-R_1\left(\bnu\sqrt{1\!+\!\be^2}\right)}
  \left(\frac{|q_{1x}|}{Q_{1}^{\NP}}\right)^{\nu C_1\cp} \\&\simeq
  \int_{-\infty}^{\infty}
  \frac{d\be\,e^{-R_1\left(\bnu\sqrt{1\!+\!\be^2}\right)}}{\pi(1+\be^2)}
  \left\{1-\nu\cp C_{1}\, \left[\> \ln( \bnu
      Q_{1}^{\NP})\>+\>\chi(\be)\>\right] \right\},
\end{split}
\end{equation}
with corrections of the second order in $\cp$.
From \eqref{eq:sigmapiccolo} and \eqref{eq:R-measureNP} we obtain the
Mellin moment
\begin{equation}
  \label{eq:R-sigma}
\sigma(\nu)\!\simeq\! 
\sigma^{\PT}(\nu)-\nu\cp\,C_1\, f_1(\nu)\>,
\qquad
f_{1}(\nu)=\sigma^{\PT}(\nu)\cdot\ln(\bnu Q_1^{\NP})\>+\>\tf_1(\nu)\,,
\end{equation}
where $\tf_1$ is given by
\begin{equation}
  \label{eq:tf1R}
\begin{split}
  \tf_1(\nu)=\int_{-\infty}^{\infty}\frac{d\be\,\chi(\be)}{\pi(1+\be^2)}\>
  e^{-R_1\left(\bnu\sqrt{1+\be^2}\right)}
  =e^{-R_1(\bnu)}\left\{E_1(\bnu)+\cC_1(\nu)\right\}.
\end{split}
\end{equation}
This is obtained by splitting the $\chi(\be)$ function (see
\eqref{eq:chi}). The function $E_1(\bnu)$ is given in \eqref{eq:E1}
(with the PT scale $Q_1^{\PT}$ of the $\Ko^R$ case), and $\cC_1(\nu)$
is the SL function
\begin{equation}
  \label{eq:R-tf1}
\cC_1(\nu)\simeq \int_{-\infty}^{\infty}
\frac{d\be\,\chi'(\be)}{\pi}
\left(\frac{1}{1+\be^2}\right)^{1+\half C_1\,r'(\nu,Q)}.
\end{equation}
Here we used the expansion
$R_1(\bnu\sqrt{1\!+\!\be^2}\,)-R_1(\bnu)\simeq
\ln\sqrt{1\!+\!\be^2}\,C_1\,r'(\nu,Q)$ since the $\be$-integration is
fastly convergent.  We now analyze the NP corrections to the mean and
the distribution.

\paragraph{Correction to the mean $\Ko^R$.}
The calculation is similar to that for the previous case. From
\eqref{eq:mean}, \eqref{eq:R-sigma} and \eqref{eq:tf1R} we obtain
\begin{equation}
  \label{eq:R-mean}
\VEV{\Ko^R}^{\NP}_{\conf}=\cp\,C_1\,f_1(0)\simeq
\cp\,C_1\VEV{\ln \frac{Q_1^{\NP}}{|q_{1x}|}}\>,
\end{equation}
with corrections of order $\sqrt{\as}$. The average value here is computed
over the parton $\#1$ distribution given in \eqref{eq:D1} (apart from
the hard scale being the one for the $\Ko^R$ case).

\paragraph{Correction to the $\Ko^R$ distribution.}
As before, from the Mellin moment \eqref{eq:R-sigma} we obtain the NP
correction to the distribution
\begin{equation}
  \de \Sigma(\Ko)=-\cp\,C_a\,\psi_1(\Ko)\>,
\qquad
\psi_1(\Ko)=\partial_{\Ko}\int\frac{d\nu}{2\pi i \nu}e^{\nu\Ko}\>f_1(\nu)\>.
\end{equation}
To evaluate $\psi_1$ we use the operator identity 
\begin{equation}
\label{eq:Opera1}
e^{-R_1\left(\bnu\sqrt{1+\be^2}\right)}
= e^{-R_1\left(e^{-\partial_z}\right)}\>
\left. \left(\bnu\sqrt{1+\be^2}\,\right)^{-z}\right|_{z=0}.
\end{equation}
to get
\begin{equation}
  \label{eq:R-psi}
\begin{split}  
\!\!\!\psi_1
\!=\! e^{-R_1\left(e^{-\partial_z}\right)}\!
\int_{-\infty}^{\infty}\!
&\frac{d\be}{\pi}\!\left(\frac{1}{1\!+\!\be^2}\right)^{1\!+\!\frac z2}\! 
\!\!\left(\!\ln\frac{Q_1^{\NP}}{\bKo}\!+\!\psi(1\!+\!z)\!-\!
\frac{1}{z}\!+\!\chi(\be)\!\right)\!
\left. \frac{\partial_{\Ko}\,\bKo^{z}}{\Gamma(1\!+\!z)}\right|_{z=0}
\!\!\!\!.
\end{split}
\end{equation}
The integration over $\chi(\be)$ generates a singular $1/z$ term:
\begin{equation}
\label{eq:fint}
\begin{split}
\!\!\int_{-\infty}^{\infty}\!\frac{d\be\,\chi(\be)}{\pi}\!
\left(\!\frac{1}{1\!+\!\be^2}\!\right)^{1\!+\!\half z}\!\!\!\!=\!
\left[\frac1z\!+\!\half\psi\left(\!1\!+\!\frac z2\right)\!-\!
\half\psi\left(\frac{\!1\!+\!z\!}{2}\right) \right]\!
\int_{-\infty}^{\infty}\!\frac{d\be}{\pi}
\left(\frac{1}{1\!+\!\be^2}\right)^{1\!+\!\half z}\!\!\!\!.
\end{split}
\end{equation}
This means that inside the integrand of \eqref{eq:R-psi} we can
replace
\begin{equation}
  \label{eq:cancellation}
-\frac1z+\chi(\be)\>\Rightarrow\>
\half\psi\left(1+\frac z2\right)-\half\psi\left(\frac{1+z}{2}\right).
\end{equation}
We conclude that the singular $1/z$ term in \eqref{eq:R-psi} is
cancelled by the contribution from the integral of the $\chi(\be)$
function in \eqref{eq:fint}.
 
Performing the $\be$-integral in \eqref{eq:R-psi} we obtain the PT
distribution ($z$ is substituted with $R_1'$).  As for the $\Ko^T$
case, the NP correction can be expressed as a shift
\begin{equation}
  \label{eq:R-shift}
\begin{split}
  \Sigma(\Ko) \>\simeq \>\Sigma^{\PT}(\Ko\!+\!\delta\Ko)\>, \qquad
  \delta\Ko=-\cp\,C_1\,\Delta_1(\Ko)\>,
\end{split}
\end{equation}
where
\begin{equation}
  \label{eq:dKoR}
\begin{split}
\Delta_1(\Ko)&=\ln \frac{Q^{\NP}_1}{\bKo} \!+\!\psi(1\!+\!R_1') 
\!+\!\half\psi\left(1\!+\!\frac{R_1'}{2}\right)
\!-\!\half\psi\left(\frac{1\!+\!R_1'}{2}\right)\\& 
=\ln \frac{2Q^{\NP}_1}{\Ko}\>\left(1+\cO{{\as}}\right).
\end{split}
\end{equation}
We see that in this case we do not have any contribution from the
singular $E_1$-function. This is due to the fact that in
\eqref{eq:cancellation} the $1/z$ singularity was exactly cancelled by
the integration over $\chi$-function.  In the $\Ko^T$ case the
corresponding cancellation was only partial.

The physical reason for this cancellation can be understood as
follows.  For the $\Ko^R$ distribution we are considering emissions
only in the right hemisphere. In this case the parton \#1 always
recoils with $|q_{1x}|\sim \Ko$, be it the regime of well-developed PT
cascades, $\as\ln^2Q/\Ko\gg1$, or the regime with few secondary
partons, $\as\ln^2Q/\Ko\ll1$.  (In this latter case even the finite
factor $2$ in the scale can be explained from the event plane
kinematics \eqref{eq:I1} and Fig.~\ref{fig:plane}.)

\section{Summary of relevant formulae and discussion \label{sec:Discuss}}
In this paper we have computed the leading NP correction to the
out-of-plane momentum distribution
defined in \eqref{eq:sigdef}
in the kinematical region \eqref{eq:TTM}.  
We have studied the distributions in the variables $\Ko^T$ (see
\eqref{eq:KodefT}) and $\Ko^R$ (see \eqref{eq:KodefR}). The NP results
obtained here, together with the SL-resummed PT calculations carried
out in \cite{acopt}, extend the accuracy of QCD description of CIS jet
observables in $\ee$ annihilation processes beyond two-jet physics.

In spite of the relative complexity of three-jet ensembles, both PT
and NP expressions have a clear physical interpretation based on
simple QCD considerations and kinematical relations.
One should expect that the present methods can be extended 
to other processes and CIS distributions in multi-jet environment.

We have analysed the NP corrections to the logarithmic ``soft factor''
$\Sigma(\Ko)$ in the general QCD expression \eqref{eq:Sig-factorized}
based on the QCD factorization of soft radiation. 
The corresponding non-logarithmic coefficient function is computed in
\cite{acoagg}.

The NP corrections have been obtained by using the dispersive method
of Ref.~\cite{DMW} for defining the running coupling and triggering
the NP effects in CIS observables.  The characteristic NP parameter
$\cp$ defined in \eqref{eq:cp} can be related, after merging of PT and
NP contributions, with the integral of the running coupling over the
infrared region. It is the same parameter that enters the NP
corrections to distributions and means of two-jet observables such as
thrust $T$, invariant jet masses $M^2$, the $C$-parameter and jet
broadenings $B_T,B_W$.  In \eqref{eq:thrust} $\cp$ is related to the
well-known NP shift in the $1\!-\!T$ distribution, see \cite{DLMS}.

The previous sections are full of formulae, so it may be helpful to
single out the important ones and to list their main physical
properties.  Before turning to the NP corrections, we recall, for the
sake of completeness, the key features of SL-resummed PT distribution
$\Sigma^{\PT}(\Ko)$.

\subsection{PT results}
We discuss first the $\Ko^T$ case and then $\Ko^R$.

\paragraph{$\Ko^T$ case.}
The PT result is given in \eqref{eq:T-SigPT} with the following structures:
\begin{itemize}
\item the three radiators $R_a(\bKo^{-1})$ in \eqref{eq:r} collect all
  DL ($\as^n\ln^{n+1}Q/\Ko$) contributions and essential SL
  corrections ($\as^n\ln^nQ/\Ko$).  $R_a$ is the standard radiator
  that one encounters in the two-jet physics \cite{PTstandards},
  slightly modified here by the factor 2 in the running coupling
  scale, which is due to the fact that the relative transverse
  momentum in the argument of the coupling is integrated over one
  (in-plane) component;
\item the hard scales $Q_a^{\PT}$ are given in \eqref{eq:Qa} and
  \eqref{eq:QPT}.  Their precise expressions are essential only in the
  arguments of the DL terms
  (the SL correction terms do not distinguish between $Q$ and
  $Q_a^{\PT}$).  It is important that these scales 
  depend on the geometry of the event (the angles between jets);
\item the additional ``recoil'' factor $\cF_T(\Ko)$ is a SL function
  (a function of $r'\propto \as\ln (Q/\Ko)$) given in \cite{acopt} as
  a multiple integral over Fourier variables. This function depends on
  the configuration $\conf$ of the underlying hard process (Born
  parton system), as can be seen from its first-order expansion
  \eqref{eq:cFTone}.  The latter expression has a simple physical
  interpretation and can be easily reconstructed from event-plane
  kinematics with a single secondary gluon (see the discussion in
  subsection~\ref{sec:recollationT}).
\end{itemize}

\paragraph{$\Ko^R$ case.}
The PT result is given in \eqref{eq:R-SigPT2} and \eqref{eq:R-SigPT}.
Here only one radiator contributes, $R_1(\bKo^{-1})$ of the most
energetic parton in the right hemisphere, so that the distribution in
$\Ko^{R}$ is sensitive to the Born configuration already at DL level.
Radiation from the other two hard parton contributes only at SL level
and can therefore be included in the hard momentum scale $Q_1^{\PT}$,
which has been computed in \cite{acopt} and does not have a simple
geometrical interpretation (see the argument given in
subsection~\ref{sec:rad-R}).
As in the $\Ko^T$ case, the first-order expression for the
distribution can be reconstructed by simply examining event-plane
kinematics.

\subsection{NP result}
We have studied the NP contribution leading in $\cp$. This
approximation is valid in the region $\LQCD\ll \Ko\ll Q$.  For values
of $\Ko$ of the order of $\LQCD$ the simple shift approximation fails,
see \cite{KS}.

We start again from the case of the $\Ko^T$ distribution.

\paragraph{$\Ko^T$ case.} 
The distribution $\Sigma(\Ko)$ with account of the leading NP
correction, is given in \eqref{eq:T-shift}. The NP correction enters
as a shift of the argument of the PT distribution, $\de\Ko$ given in
\eqref{eq:dKoT}.  The NP gluon emitted by the hard parton $\#a$
contributes to the shift with two terms $\Delta_a^{(1)}(\Ko)$ and
$\Delta_a^{(2)}(\Ko)$ given in \eqref{eq:T-Delta1} and
\eqref{eq:Delta-2} respectively.  Corrections to these expressions are
of the relative order $\as$ and are beyond our accuracy.

The first term $\Delta_a^{(1)}(\Ko)$ is additive (independent
emission) since its contribution to $\de\Ko$ is proportional to the
colour charge $C_a$ of the emitter.  The functions
$\Delta_a^{(1)}(\Ko)$ behave logarithmically in $\Ko$, with the hard
NP scales $Q_a^{\NP}$ given in \eqref{eq:QNP} and \eqref{eq:Qa}. These
scales differ from the corresponding PT scales \eqref{eq:QPT} only by
constant factors ($\zeta^{\NP}$ versus $\zeta_a^{\PT}$), which are due
to small-angle radiation effects.

The second term $\Delta_a^{(2)}(\Ko)$ is expressed in terms of the PT
functions $\cF_T(\Ko^{-1})$, $\cF_a(\Ko^{-1})$ and $\cC_a(\Ko^{-1})$,
and the NP function $E_a(\Ko^{-1})$.  In particular, $\cF_T(\Ko^{-1})$
and $\cF_a(\Ko^{-1})$ are the SL functions encountered in the PT
analysis.  The first is given in \cite{acopt} and the second in
\eqref{eq:cFa}.  The SL functions $\cC_a(\Ko^{-1})$ are defined in
Appendix~\ref{App:tfa}.  Finally, the NP functions $E_a(\Ko^{-1})$,
defined in \eqref{eq:E1} and \eqref{eq:E2}, are typical for
observables that are uniform in the rapidity of the emitted particles.
They have first emerged in the study on the NP correction to the total
broadening distributions in \cite{broad}.

One has the following two regimes for the shift:
\begin{itemize}
\item for very small $\Ko$ ($\as\ln^2Q/\Ko\gg1$), when the QCD cascade
  is well-developed and there are many secondary partons around, the
  shift $\de\Ko$ is dominated by $\Delta_a^{(1)}$, is logarithmic in
  $\Ko$ and proportional to the total colour charge $C_T=2C_F+N_c$ of
  the three-parton system:
\begin{equation}
  \label{eq:uno}
  \de\Ko\simeq -\cp\,C_T\,\ln\frac{Q}{\Ko}\>.
\end{equation}
In this region the NP radiation effects are additive;
\item in the opposite region with moderately small $\Ko$
  ($\as\ln^2Q/\Ko\ll1$), when there are few secondary partons, the
  shift is dominated by the contributions $\Delta_a^{(2)}$ that are
  singular in $\as$, and the dependence on the colour charges becomes
  rather odd:
\begin{equation}
  \label{eq:due}
  \de\Ko\simeq -\frac{\pi\,\cp}{2 C_T\sqrt{\as}}\left(
\frac{C_1(C_2+C_3)}{\sqrt{C_1}}
+\frac{C_2C_3}{\sqrt{C_1+C_2}}
+\frac{C_2C_3}{\sqrt{C_1+C_3}}\right).
\end{equation}
\end{itemize}
This peculiar colour structure results from an interplay between the
NP and PT radiation effects.  Its physical origin is discussed in
detail in subsection~\ref{subsec:distribution}, below
\eqref{eq:Delta}.

\paragraph{$\Ko^R$ case.} 
The $\Ko^R$ distribution, with account of the leading NP correction,
is given in \eqref{eq:R-shift} with the NP shift given in
\eqref{eq:dKoR}.  The procedure for computing the hard NP scale
$Q_1^{\NP}$ is presented in Appendix~\ref{App:RadNP}.  This scale does
not have a simple geometrical meaning.  In agreement with the usual
argument based on the event plane kinematics (see \eqref{eq:I1} and
the discussion after \eqref{eq:dKoR}), the shift in the $\Ko^R$
distribution is purely logarithmic in $\Ko$. Unlike the $\Ko^T$ case,
it does not contain singular $1/\sqrt{\as}$ contributions.  This
feature is similar to that of the single-jet broadening distribution
\cite{broad}.

\paragraph{NP corrections to the $\Ko$ means.}
The results are given in \eqref{eq:T-mean2} and \eqref{eq:R-mean} for
the total and the right-hemisphere observables, respectively. They are
expressed in terms of the unrestricted averages \eqref{eq:VEV1} and
\eqref{eq:VEV23}.  Again, the peculiar colour-charge dependence is due
to the PT--NP interplay.

\subsection{Final considerations}

To demonstrate the structure of $\Ko$ spectra under study we show in
Fig.~\ref{fig:acot} and Fig.~\ref{fig:acor} distributions generated by
Herwig \cite{herwig} for the three Born configurations.  Notice that
the difference between the distributions for each configuration is
much more evident in the right case since it appears already at DL
level.  The corresponding numerical predictions following from the
present theoretical analysis, including the relative weight of PT and
NP components, will be presented elsewhere \cite{acoagg}.

We are intended to explore whether the approach we have developed here
for the analysis of the PT and NP contributions to the $\Ko$
distributions and means can be extended to more complicated processes
and other multi-jet observables.
What makes one hopeful is the observation that, in spite of technical
complications, the final results, both in the PT and NP sectors, do
possess a simple physical interpretation.

In particular, we plan to analyze the case with one or two jets in the
initial state (lepton-hadron and hadron-hadron collisions,
respectively). The resulting distribution should be sensitive to the
structure of the underlying hard matrix element, as it is the case for
the distributions considered in this paper.
Carrying out such a programme, and comparing the QCD predictions with
experiment, should provide a deeper understanding of the hadronization
physics.

\paragraph{Acknowledgements}

We are grateful to Gavin Salam for helpful discussions and suggestions.


\EPSFIGURE[ht]{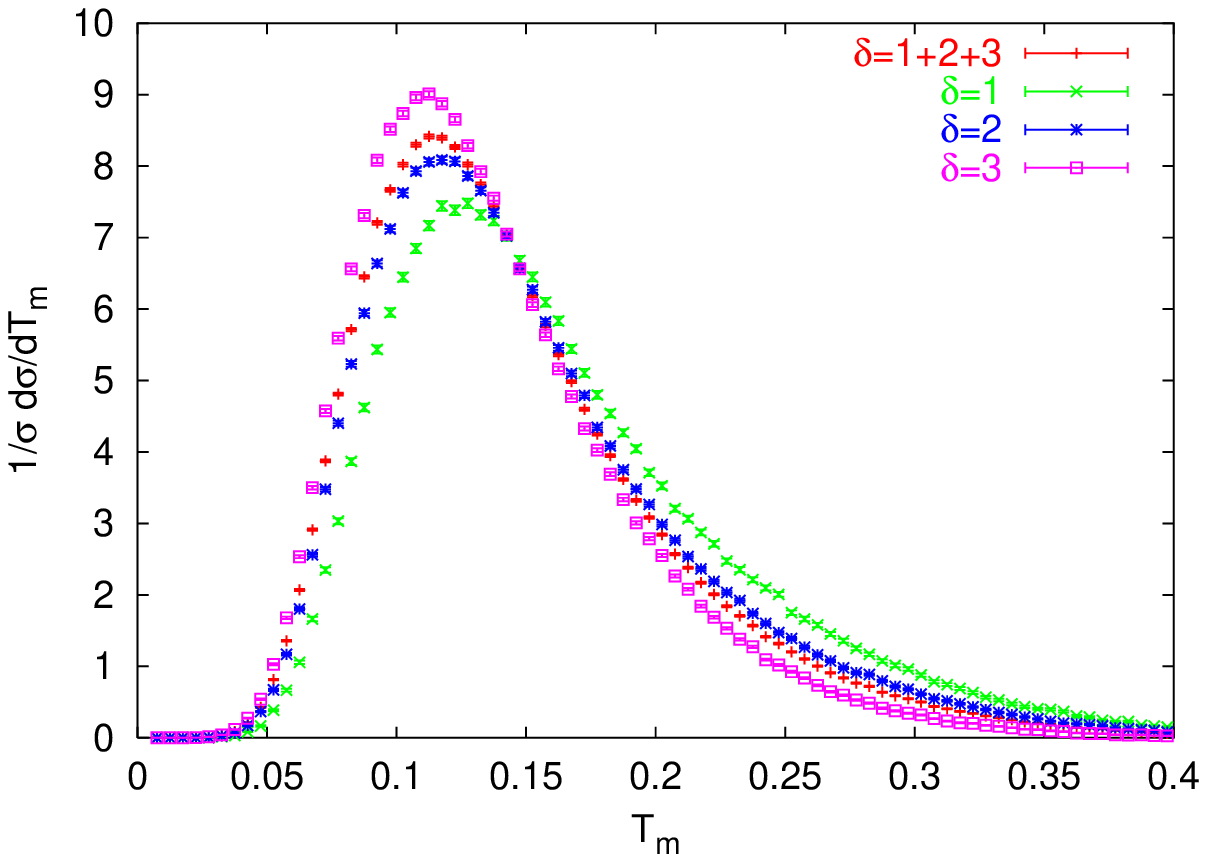,width=1.0\textwidth}{Herwig simulation for
  the total minor ($T_m=\Ko^T/Q$) distribution with
  \mbox{$y_{cut}=0.1$} (Durham algorithm \cite{durham}). The picture
  shows the distribution for the three underlying configurations and
  their weighted sum. \label{fig:acot}}

\EPSFIGURE[ht]{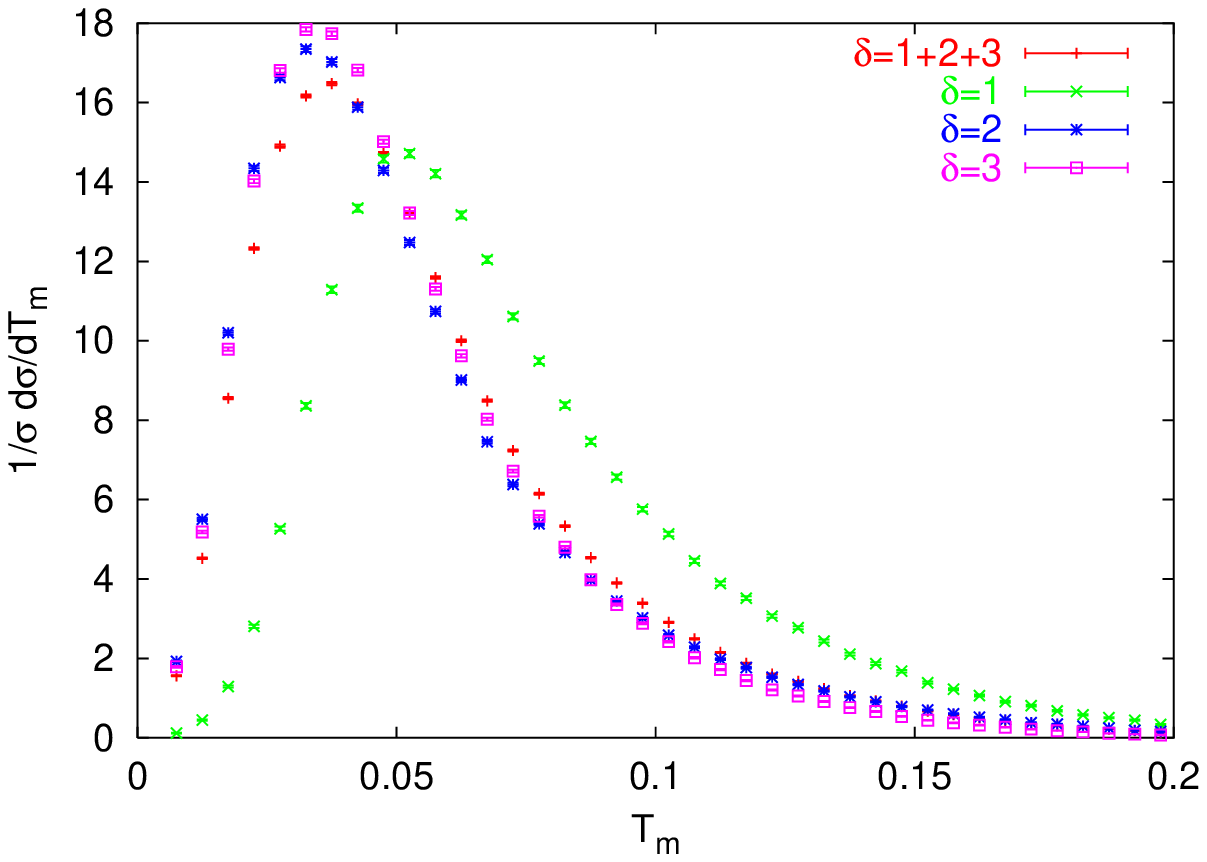,width=1.0\textwidth}{Herwig simulation for the
  narrow minor ($T_m=\Ko^R/Q$) distribution with \mbox{$y_{cut}=0.1$} (Durham
  algorithm). The picture shows the distribution for the three
  underlying configurations and their weighted sum. \label{fig:acor}}

\newpage
\appendix

\section{Soft radiation resummation and PT distributions\label{App:RadPT}}
The resummation of the PT contributions to the distribution
$\Sigma^{T/R}_{\conf}$ to SL accuracy has been performed in
\cite{acopt}.  To achieve such an accuracy one needs $M_{n,\conf}^2$
at two-loop level. The recoil momenta $q_a$ need to be taken into
account in the kinematics, but they can be neglected in
$M_{n,\conf}^2$.  The essential point for the resummation of the soft
radiation is the factorization of $M_{n,\conf}^2$ and of $dH_n^{R/T}$,
the soft multi-parton distributions and the phase spaces
\eqref{eq:dHnT} and \eqref{eq:dHnR}.  In this Appendix we recall the
PT relevant points needed to analyze the NP corrections in
$\Sigma(\Ko)$ which emerge as an interplay between PT and PT
contributions.

Before turning to $\Sigma^{\PT}(\Ko)$ we recall the structure of the
soft matrix element factorization at two-loop level.  We start with
the one loop expression, without virtual corrections, given by
\begin{equation}
\label{eq:Mn2f}
  M_{n,\conf}^2 \>\simeq \> \prod_{i=1}^n W_{\conf}(k_i)\>,
\end{equation}
where $W_{\conf}$ is the distribution of soft gluon radiation off the
hard three-parton antenna in the configuration ${\conf}$. For instance
for $\conf=3$ one has
\begin{equation}
  \label{eq:W}
\begin{split}
W_{3}(k)&=\frac{N_c}{2}\left(w_{13}+w_{23}-\frac{1}{N_c^2}w_{12}\right),\\
w_{ab}(k)&=\frac{\as} {\pi\,k^2_{ab,\,t}} \>,
\qquad k^2_{ab,\,t}= \frac{2(p_ak)(kp_b)}{p_ap_b}\>,
\end{split}
\end{equation}
with $k_{ab,\,t}$ the invariant transverse momentum of $k$ with
respect to the $ab$-dipole. In general we can write
\begin{equation}
  \label{eq:W'}
W_{\conf}(k)=\sum_{a<b}C_{ab}^{(\conf)}w_{ab}(k)\>,
\quad C_{ab}^{(a)}=C_{ab}^{(b)}=\frac{N_c}{2}\>, 
\quad C_{ab}^{(c)}=\frac{-1}{2N_c}\>,\>\> c \ne a,b\,.
\end{equation}
This expression can be generalized at two-loop by considering the
emission of a pair of soft partons from the three-jet configuration
\cite{2soft}.  The two-soft emission distribution has the same form as
\eqref{eq:W} and this justifies the exponentiation of the two-loop
soft emission distribution. As shown in \cite{DLMS}, the two loop
corrections to the soft matrix element enter at PT level simply by
properly setting the argument of the running coupling in \eqref{eq:W}
at the dipole invariant transverse momentum.  Since for the PT
analysis at SL accuracy we can neglect the recoil in the soft matrix
element, we set $p_a=P_a$ in $w_{ab}$.

The factorization of the phase space $dH^{T/R}_n$ is achieved by using
the Fourier and Mellin representations for the delta and theta
functions in \eqref{eq:dHnT} or \eqref{eq:dHnR}. Each soft parton
$k_i$ enters with a source $u(k_i)$ which, at PT level, must satisfy
the condition $u(k_i)\!\to\! 1$ for $k_i$ collinear to $P_a$.  In the
next two sections we recall the form of the sources, the radiators and
the PT distributions for the $\Ko^T$ and $\Ko^R$ cases in succession.

\subsection{$\Ko^T$ case: source and PT radiator \label{App:sourceT}}
We use
\begin{equation}
  \label{eq:mfT}
\begin{split}
  \vartheta\left(\!\Ko\!-\!\sum_{a=1}^3\abs{q_{ax}}\!-\!
    \sum_{i=1}^n\abs{k_{ix}}\!\right)\>=&
  \>\int\frac{d\nu\,e^{\nu\Ko}}{2\pi i\nu}\>
  \prod_{a=1}^3 e^{-\nu|q_{ax}|}\> \prod_i e^{-\nu|k_{ix}|}\>,\\
  \delta^2\left(\vec{q}_{1}+\sum_R\vec{k}_{it}\right) \>=&\>
  \int\frac{\nu^2\,d^2b_1\,e^{i\nu\vec{b_1}\cdot\vec{q}_{1}}}{(2\pi)^2}\>
  \prod_R e^{i\nu\vec{b_1}\cdot \vec{k}_{it}}\>, \\
  \delta\left(q_{2x}+q^+_{1x}+\sum_{U}k_{ix}\right)\>=&\>
  \int\frac{\nu\,d\be_2\,e^{i\nu\be_2\left(q_{2x}+q^+_{1x}\right)}}{2\pi}
  \> \prod_{U}e^{i\nu\be_2k_{ix}}\>,\\
  \delta\left(q_{3x}+q^-_{1x}+\sum_{D}k_{ix}\right)\>=&\>
  \int\frac{\nu\,d\be_3\,e^{i\nu\be_3\left(q_{3x}+q^-_{1x}\right)}}{2\pi}
  \>\prod_{D}e^{i\nu\be_3 k_{ix}}\>,
\end{split}
\end{equation}
with $q^{\pm}_{1x} \equiv q_{1x}\> \vartheta(\pm q_{1y})$.  We recall
that $R=C_1+C_4$, $U=C_1+C_2$ and $D=C_3+C_4$ represent the right-,
up- and down-regions of the phase space with $C_\ell$ the regions in
the four quadrants.  Each emitted soft parton contributes with the
source (for the PT case)
\begin{equation}
  \label{eq:u0T}
u_T^{(0)}(k)=\sum_{\ell=1}^4 \Theta_\ell(k)\>u^{(0)}_\ell(k)\>,
\end{equation}
where $\Theta_\ell(k)$ is the support function for $k\in C_\ell$ and
\begin{equation}
  \label{eq:u0ell}
\begin{split}
  &u^{(0)}_1(k)=e^{-\nu\left\{ |k_x|-i\be_{12}k_x -i\gam
      k_{y}\right\}}\,,\qquad
  u^{(0)}_4(k)=e^{-\nu\left\{ |k_x|-i\be_{13}k_x -i\gam k_{y}\right\}}\,,\\
  &u^{(0)}_2(k)=e^{-\nu\left\{ |k_x|-i\be_{2}k_x \right\}}\,,
\qquad\qquad\>
  u^{(0)}_3(k)=e^{-\nu\left\{ |k_x|-i\be_{3}k_x \right\}}\,,\\
&  
\be_1=b_{1x}\>,\quad  \gam=b_{1y}\>,\quad
\be_{12}\equiv \be_{1}+\be_2\>,\quad 
\be_{13}\equiv \be_{1}+\be_3\>.
\end{split}
\end{equation}
The source $u_T^{(0)}(k)$ depends on $k_x$ and $k_y$ for $k\in C_1$ or
$C_4$, and only on $k_x$ for $k\in C_2$ or $C_3$. We obtain
\eqref{eq:dhn} and \eqref{eq:dmu}, where
\begin{equation}
  \label{eq:VT}
\begin{split}
V^T(\gam\be q)\>=\>
&e^{-\nu(|q_{2x}|-i\be_2 q_{2x})}\,
e^{-\nu(|q_{3x}|-i\be_3 q_{3x})}\,
e^{i\nu\gam q_{1y}}
\\&\times
\left[
 \vartheta( q_{1y})\, e^{-\nu(|q_{1x}|-i\be_{12} q_{1x})}\,+\,
 \vartheta(-q_{1y})\,e^{-\nu(|q_{1x}|-i\be_{13} q_{1x})} \right].
\end{split}
\end{equation}
The source \eqref{eq:u0T} satisfies the condition $u^{(0)}_T(0)=1$ and
therefore constitutes a suitable source for the PT analysis at SL
accuracy where recoil can be neglected.  

By using the factorization of soft gluon emission \eqref{eq:Mn2f} and
of the phase space \eqref{eq:dhn} we can resum 
\eqref{eq:Sigdef} and obtain
\begin{equation}
  \label{eq:fnt1}
\begin{split}
\sum_n\frac{1}{n!}\int dH_n^T\,\prod_{i}^n [dk_i]\,M_{n,\conf}^2
&\>\Rightarrow\>
\sum_{n=0}^{\infty} \frac{1}{n!}
\int \prod_i [dk_i]\, W_{\conf}(k_i)\,u_T^{(0)}(k_i)
\\&\>=\>\exp\left\{\int [dk]\,W_{\conf}(k)\,u_T^{(0)}(k)\right\},
\quad [dk]=\frac{d^3k}{\pi\omega}
\end{split}
\end{equation}
where, in the r.h.s we removed the integration $d\mu^T$, see
\eqref{eq:dmu}.

Next we include the virtual corrections by subtracting unity to the
source and obtain the PT radiator for the total distribution
\begin{equation}
  \label{eq:RadT}
\int[dk]\,W_{\conf}(k)\,u_T^{(0)}(k)
\>\Rightarrow\>
-\int [dk]\,W_{\conf}(k)\,[1-u_T^{(0)}(k)]=
-\cR^{\PT}_{\conf}(\nu,\be_a,\gam)\>.
\end{equation}
This expression for the PT radiator of $\Sigma^{\PT}$ 
\begin{equation}
\label{eq:rabT}
\cR^{\PT}_{\conf}(\nu,\be_a,\gam)
=\sum_{a<b}C_{ab}^{(\conf)}\>r_{ab}\>,
\qquad 
r_{ab}= \int \frac{d^3k}{\pi\om}\>w_{ab}(k)\>
[1-u_T^{(0)}(k_x)]\>,
\end{equation}
is valid at two-loop accuracy provided one uses the correct argument
of $\as$. One has
\begin{equation}
\label{eq:wabdef}
w_{ab}(k)=\frac{\as(k^2_{t,ab})}{\pi k^2_{t,ab}}\>,
\end{equation}
with the coupling defined in the physical scheme \cite{CMW}.  This
radiator has been evaluated in \cite{acopt} and is given in
\eqref{eq:T-RadPT}.

The Mellin moment $\sigma^{\PT}_{\conf}(\nu)$ is given by the $d\mu^T$
integral over the three ``Sudakov form factors'' associated to the
three hard partons.  Since the PT radiator does not depend on the
recoil, we can freely integrate the function $V^T$ over $q_{1y}$ and
$q_{ax}$ and get
\begin{equation}
  \label{eq:T-I0}
\begin{split}  
I(\gam,\be)
&=\frac{\nu^4}{8} \int dq_{1y} \prod_adq_{ax}\>V^T(\gam\be q)\\
&=\frac{1}{(1+\be^2_2)(1+\be^2_3)}
\left(\frac{1}{(1+\be^2_{12})(-i\gam+\eps)}+
      \frac{1}{(1+\be^2_{13})( i\gam+\eps)}\right),
\end{split}
\end{equation}
To obtain the Mellin moment we need to integrate over the Fourier
variables.  Consider the integral
\begin{equation}
\label{eq:Zdef}
\begin{split} 
Z(\nu,\be_{23})\!\equiv\!\!
\int_{-\infty}^{\infty}\!\frac{d\be_1}{\pi}
\int_{-\infty}^{\infty}\!\frac{d\gam}{2\pi}\,e^{-\bR_1}
\left(\frac{1}{(1\!+\!\be^2_{12})(-i\gam\!+\!\eps)}\!+\!
      \frac{1}{(1\!+\!\be^2_{13})( i\gam\!+\!\eps)}\right),
  \end{split}
\end{equation}
with $\bR_1=\bR_1(\bnu,\be_{12},\be_{13},\gam)$ given in
\eqref{eq:T-RadPT}.  The symmetry of the $\be_a,\gam$-integrations
implies that $Z$ is a function of the sum $\be_{23}=\be_2+\be_3$.  The
radiator $\bR_1$ has an imaginary part; however, due to the symmetry
\begin{equation}
  \label{eq:symmetry}
\bR_1(\bnu,\be_{12},\be_{13},\gam) =  
\bR_1(\bnu,\be_{13},\be_{12},-\gam)=  
\left\{\bR_1(\bnu,\be_{13},\be_{12},\gam)\right\}^*,
\end{equation}
and to the properties of the $\be_2,\be_3$-integrations, the final result is 
real and reads
\begin{equation}
\label{eq:Z}
\begin{split}
Z(\nu,\be_{23})=\>\int_{-\infty}^{\infty}
\frac{d\be}{\pi(1+\be^2)}\>{\cS}(\nu,\be, \be_{23})\>,
\end{split}
\end{equation}
with $\cS$ given in \eqref{eq:cS}. 
The real and imaginary parts of $\bR_1$ are given, respectively, by
\begin{equation}
\label{eq:T-Rad1-ri}
\begin{split}
&\bR_1^{(r)}(\bnu,\be,\be',\gam)= 
\half R_1(\bmu)+\half R_1(\bmu')\>,\\
&\bR_1^{(i)}(\bnu,\be,\be',\gam)= 
R'_1(\bmu)\,B(\be,\gam)-R'_1(\bmu')\,B(\be',\gam)\>,
\end{split}
\end{equation}
where
\begin{equation}
  \label{eq:bmu}
\bmu\equiv\bnu\sqrt{(1+ \gam)^2+\be^2}\>, 
\qquad
\bmu'\equiv\bnu\sqrt{(1+ \gam)^2+{\be'}^2}\>,
\end{equation}
and where the function $B(\be,\gam)$ is given by
\begin{equation}
  \label{eq:B}
\begin{split}
B(\be,\gam)&\equiv
\int_0^\infty\frac{dx}{2\pi(1-x^2)} 
 \ln\frac{(1+x\gam)^2+\be^2}{(1+ \gam)^2+\be^2}
\\&=
\int_0^1\frac{dx}{2\pi(1-x^2)} 
 \ln x^2\frac{(1+x\gam)^2+\be^2}{(x+ \gam)^2+x^2\be^2}\>.
\end{split}
\end{equation} 
This function has the following limits 
\begin{equation}  
  \label{eq:Blim}
\begin{split}
B(\be,\gam)\>\to\> 0             \>\>\mbox{for}\>\>1,\gam\ll\be,\qquad
B(\be,\gam)\>\to\> -\frac{\pi}{4}\>\>\mbox{for}\>\>1,\be\ll\gam\>.
\end{split}
\end{equation}

\subsection{$\Ko^R$ case: source and PT radiator \label{App:sourceR}}
We factorize the phase space by using Mellin and Fourier
representation and we obtain \eqref{eq:dhn} and \eqref{eq:dmu} with
\begin{equation}
  \label{eq:VR}
V^R(\be,q_{1x})=\,e^{-\nu(|q_{1x}|-iq_{1x}\be)}\>,
\end{equation}
and the source for the soft parton emission 
\begin{equation}
  \label{eq:u0}
  u_R^{(0)}(k_x)=u^{(0)}(k)\>\Theta_R(k)+\Theta_L(k)\,,
\qquad u^{(0)}(k) \equiv e^{-\nu(|k_x|\,-\,i\be k_x)}\>,
\end{equation}
where $\Theta_{R/L}(k)$ is the support function for $k\in R/L$.  This
source satisfies the condition $u^{(0)}(0)=1$ and therefore is a
suitable source for the PT analysis at SL accuracy.

Proceeding as before, to SL accuracy, the PT radiator for the $\Ko^R$
distribution is given by
\begin{equation}
  \label{eq:rabR}
\cR_{\conf}^{\PT}=\sum_{a<b}C_{ab}^{(\conf)}\,r_{ab}\,, \qquad
r_{ab}= \int \frac{d^3k}{\pi\om}\>w_{ab}(k)\>
[1-u_R^{(0)}(k)]\>,
\end{equation}
$w_{ab}(k)$ given in \eqref{eq:wabdef}.  To SL accuracy the
$\Ko^R$-radiator is given by \eqref{eq:R-RadPT} with $Q_1^{\PT}$
computed in \cite{acopt}\footnote{This scale depends on the
  configuration index $\conf$.  In this reference we denote this scale
  for the configuration $\conf$ as $Q_{\conf}^{\PT}$}.

\section{NP correction to the radiators\label{App:RadNP}}
In this appendix we compute the NP correction to the radiators for the
$\Ko^{R/T}$ distributions. We follow the method used in \cite{DLMS} to
compute NP corrections to various collinear and infrared safe
observables in two-jet events.

We start by recalling the relevant points in the calculation of the NP
corrections to the radiators for the two-jet CIS observables which we
need to generalize to the present analysis. We focus our attention on
the broadening observable \cite{broad} which takes contributions from
the transverse momentum of the emitted particles.

\subsection{Recollection of two-jet case \label{App:2jet}}

\paragraph{Kinematics.}  
One introduces two opposite Sudakov momenta $P$ and $\bar P$, with $P$
along the thrust axis. For a (secondary) parton $k$ of mass $m$ the
Sudakov variables are given by
\begin{equation}
  \label{eq:Sud}
  \begin{split}
&    \!\!       P\!=\!\frac{Q}{2}(1,0,0,1)\>,
\quad \bar P\!=\!\frac{Q}{2}(1,0,0,-1)\,,
\quad 
k\!=\!\al P\!+\!\be\bar P+k_{t}\,, 
\quad \al\be\!=\!\frac{{k}^2_{t}\!+\!m^2}{Q^2}\,,
\end{split}
\end{equation}
with $k_{t}$ orthogonal to $P$ and $\bar P$. The right region $R$ is
defined as
\begin{equation}
  \label{eq:right}
k\in R:\qquad\qquad
  \al>\be\quad\Rightarrow\quad \al>\frac{\sqrt{k^2_{t}+m^2}}{Q}\>.
\end{equation}
To compute NP corrections to the broadening one needs to take into
account the recoil of the primary quark and antiquark momenta for
which we have a similar Sudakov decomposition.

\paragraph{Splitting of the radiator in three pieces.} 
The full radiator is a CIS quantity. One rearranges \cite{DLMS} the
two-loop radiator as a sum of three CIS terms which have been called
{\it naive}, {\it inclusive}\/ and {\it non-inclusive}\/ pieces.  The
naive piece is the contribution for the emission of a single massive
gluon together with the corresponding virtual corrections.  Its form
is given by
\begin{equation}
\label{eq:2R}
   R^{2\!-\!\jet}_{\naive}=
\frac{C_F}{\pi}\int_0^Q dm^2\aef(m^2)\frac{-d}{dm^2}
   \int_0^{Q^2} \frac{d^2\ka}{\pi (\ka^2+m^2)} 
   \int_{\frac{\sqrt{\ka^2+m^2}}{Q}}^{1}\frac{d\al}{\al}\>[1-u(k)]\>,
\end{equation} 
where $\al$ is the Sudakov variable, $\vka$ the transverse momentum of
the soft massive gluon with respect to the hard emitting parton
(including the hard parton recoil).  The integration is over the right
hemisphere \eqref{eq:right} for the massive gluon. The effective
coupling $\aef(m^2)$ is given in terms of the physical coupling in
\cite{DMW} through a dispersive representation.

\paragraph{Massive source.}  
As already stated, the source $u(k)$ takes into account the recoil of
the two hard partons and is then given in term of the relative
transverse momentum $\ka$ of the massive gluon.  One is free to chose
any prescription to implement the mass dependence in the source,
provided that $u(k)$ reduces to the massless case for $m\to 0$.
Following \cite{DLMS}, we assume that the massive gluon source is
given by the expression of the massless gluon source in which we make
the replacement
\begin{equation}
    \label{eq:mass}
 \ka\to \sqrt{\ka^2+m^2}\>.
 \end{equation}
\paragraph{Milan factor and universality.}  
Using this prescription for the massive source, one has that the naive
piece is the only one that needs to be computed since the remaining
pieces (the inclusive and the non-inclusive pieces) contribute in a
universal way simply by rescaling the naive piece by a universal Milan
factor.

\paragraph{Linearization of the sources.} 
The leading NP correction coming from \eqref{eq:2R} is obtained by
taking the linear contribution in $m$ and $\ka$ of the integrand, that
is the first term in the expansion of $[1\!-\!u(k)]$.  This fact has
important implications on which are the relevant kinematical and
Fourier variables affecting the leading NP corrections.  They are the
variables which enter the expression of the linearization of
$[1\!-\!u(k)]$ and do not vanish upon the integration over the
azimuthal angle of $\vka$ in \eqref{eq:2R}.  In the case of broadening
\cite{broad} the Fourier variable $\vec{b}$, conjugated to the
momentum $\vka$ does not enter the leading NP correction to the
radiator. Its linear contribution to $[1\!-\!u(k)]$ is proportional to
$\vec{b}\cdot\vka$ which vanishes upon azimuthal integration.

\subsection{Sudakov variables and soft distribution for the $ab$-dipole}
We now return to our three-jet case and we start to discuss the
kinematical variables.
The radiator is given by the sum of three dipole contributions
$r_{ab}$ (for the PT case see for instance \eqref{eq:rabT}). To treat
each dipole contribution $r_{ab}$ by the same method used in the
two-jet case for \eqref{eq:2R} one needs to perform a Lorentz
transformation.
Consider the $e^+e^-$ laboratory system defined by the Sudakov
variables introduced in \eqref{eq:Sud} with $P$ along the thrust axis.
In this system the two massless partons $P_a,P_b$ initiating the
$ab$-dipole emission are not in opposite directions. In order to
follow the method of \cite{DLMS}, we use the fact that the
distribution $w_{ab}$ is Lorentz invariant and we analyze $r_{ab}$ in
the $ab$-dipole center of mass (c.m.). There are three dipoles and
then there are three different c.m. systems. Given a dipole $ab$, we
introduce the $ab$-dipole c.m. and express the distribution
$w_{ab}(k)$ in these variables. We then apply the two-jet case
analysis to our case. We describe the new variables.

\paragraph{Dipole $ab$ center of mass frame.} 
Consider the laboratory frame in which the Born momenta $P_a,P_b$ of
the $ab$-dipole are given in \eqref{eq:Pa}. We go from this frame to
the $ab$-dipole c.m.  frame by a rotation in the $yz$-plane, a
$z$-boost and again a rotation in the $yz$-plane which takes $P'_a$
and $P'_b$ along the $z$-axis. We denote by $P'_a,P'_b,p'_a,p'_b$ and $k'_i$ 
the parton momenta in this frame with Sudakov components 
\begin{equation}
  \label{eq:Sab}
  \begin{split}
&     P'_a=\frac{Q_{ab}}{2}(1,0,0, 1)\,,
\quad P'_b=\frac{Q_{ab}}{2}(1,0,0,-1)\,,
\quad Q_{ab}^2=2P_aP_b\>,\\
&p'_a=A'_{ab}P'_a+b'_{ab}P'_b+q'_{at}\,,\quad A'_{ab}\simeq1\\
&p'_b=a'_{ba}P'_a+B'_{ba}P'_b+q'_{bt}\,,\quad B'_{ba}\simeq1\\
&k'_i=\al'_iP'_a+\be'_iP'_b+k'_{it}\,,\> 
\qquad \al'_i\be'_i=\frac{{k'}^2_{it}}{Q^2_{ab}}\>.
\end{split}
\end{equation}
The transverse components are orthogonal to $P'_a$ and $P'_b$. All
Sudakov variables are soft except for $A'_{ab}$ and $B'_{ba}$ which
are close to one.  The $x$-components are not affected by the Lorentz
transformation which is in the $yz$-plane, so that the components
entering our observable $\Ko$ are the same in both frames
\begin{equation}
  \label{eq:xcomp}
 q'_{ax}=q_{ax}\>, \quad
 q'_{bx}=q_{bx}\>, \quad
 k'_{ix}=k_{ix}\>.
\end{equation}
We denote by $F_{ab}$ the {\it forward}\/ region of secondary parton
$\#i$ in the $ab$-dipole c.m.
\begin{equation}
  \label{eq:forw}
k_i\in  F_{ab}: \qquad\qquad 
\al'_i>\be'_i\>\>\Rightarrow\>\> 
\al'_i>\frac{k'_{it}}{Q_{ab}}\>.
\end{equation}
The Sudakov variables \eqref{eq:Sab} are different for different
dipoles. We do not need to add an index to specify the dipole since
these variables will be used only in the calculation of the specific
$ab$-radiator.

\paragraph{Soft distribution in the $ab$-dipole c.m.} 
The (Lorentz invariant) $ab$-dipole soft distribution $w_{ab}$ is
given in terms of $k_{ab,t}$, the ``relative transverse momentum in
the $ab$-dipole frame''.  By using the Sudakov variables \eqref{eq:Sab}
in the $ab$-dipole c.m. frame we have
\begin{equation}
  \label{eq:ktab}
  \frac{1}{k^2_{ab,t}}\equiv
\frac{{k'}_t^2}{
\left(\vec{k}'_t-\al'{\vec{q}\,'_{at}}\right)^2
\left(\vec{k}'_t-\be'{\vec{q}\,'_{bt}}\right)^2}\>,
\end{equation}
where we have neglected the large components $A'_{ab}\simeq
B'_{ba}\simeq1$.

We split the dipole distribution $w_{ab}$ into two pieces:
\begin{equation}
  \label{eq:decow}
  w_{ab}=w^a_{ab}+w^b_{ab}\>,
\end{equation}
where $w^a_{ab}$ has support only in the forward region $F_{ab}$,
\begin{equation}
  \label{eq:wab}
w^a_{ab}(k)\equiv w_{ab}(k)\cdot \theta(\al'\!-\!\be') 
\simeq \frac{\as(\ka^2)\theta(\al'\!-\!\be')}{\pi \ka^2}\,,
\qquad
\vka\equiv\vec{k}'_t-\al'{\vec{q}\,'_{at}}\,.
\end{equation}
The ``backward'' piece $w^b_{ab}$ is similarly defined.  Notice that
$w_{ab}(k)$ is invariant, but the splitting in the two regions is
performed in the $ab$-dipole c.m.

In three-jet events the soft distribution $W^{(\conf)}(k)$ is given by a
sum of dipole soft distributions $w_{ab}(k)$, see \eqref{eq:W'}.   
We decompose $w_{ab}$ into the two contributions $w_{ab}^a$ and
$w_{ab}^b$ as given in \eqref{eq:decow} and we have
\begin{equation}
\label{eq:decoW}
\begin{split}
W^{(\conf)}(k)=\sum_{(ab)}C_{ab}^{(\conf)}w^a_{ab}(k)\>,
\end{split}
\end{equation}
where the sum is extended to the six {\it ordered pair}\/ $(ab)$, i.e.\ 
$ab=12,13,21,23,31,32$.
Each term contributes only in the region $F_{ab}$ (recall that
$w^a_{ab}$ has a support only in the forward region \eqref{eq:forw}).

\subsection{Source with recoil for the $\Ko^T$ distribution 
\label{App:sourceTNP}}
As discussed above, the leading NP corrections to the radiator are
obtained from the first order expansion of $[1\!-\!u_T(k)]$ in the
momentum components. This implies that, as in the broadening case, the
leading NP radiator does not depend on the four Fourier variables
$\be_a$ and $\gam$. Their first order contributions in the expansion
of $[1\!-\!u_T(k)]$ are linear in the momentum components and then
vanish upon the symmetric phase space and Fourier variable
integration.  We have that only the Mellin variable enters the leading
NP radiator in the form $\nu|k_{ix}|$.  The conclusion is that in the
phase space $dH_n^{T/R}$ in \eqref{eq:dHnT} and \eqref{eq:dHnR} we
should take into account the NP gluon only in the theta function
giving the upper bound $\Ko^{T/R}$. On the contrary, the NP soft gluon
does not contribute to the determination of the event plane.  The
delta functions which fix the event plane should be taken into
account only in the PT contribution.

We now analyze how one can take into account in the sources the recoil
coming from the observable $\Ko^T$.  As in \eqref{eq:fnt1} we start
from the sum
\begin{equation}
  \label{eq:T-sum}
\begin{split}
\sum_n\frac{1}{n!}\!\int\! dH_n^T \prod_i[dk_i] M_{n,\conf}^2
\!=\!\int\!\frac{d\nu\,e^{\nu \Ko}}{2\pi i \nu}
\int\!d\mu^T\!\exp\left\{\int[dk]W_{\conf}(k) u_T^{(0)}(k)\!\right\}.
\end{split}
\end{equation}
Using the expression in \eqref{eq:decoW} for the soft gluon
distribution we can write
\begin{equation}
  \label{eq:T-sum1}
\begin{split}
\exp\left\{\int[dk]W_{\conf}(k)u_T^{(0)}(k)\right\}=
\prod_{(ab)}\exp\left\{C^{(\conf)}_{ab}\int\,
[dk]\,w^a_{ab}(k)\,u_T^{(0)}(k)\right\}\>.
\end{split}
\end{equation}
From \eqref{eq:wab} one has that $w^a_{ab}(k)$ is collinear singular
for
\begin{equation}
  \label{eq:ab-sing}
\vka=\vec{k}'_t-\al'\vec{q}\,'_{at}\to 0\,,
\quad\Rightarrow\quad k_x-\al'q_{ax}\to0\>,
\end{equation}
(with the prime variables defined in \eqref{eq:Sab}, the $ab$-dipole
c.m.). From the form \eqref{eq:u0T} of these source we have
$u_T^{(0)}(k)\ne1$ in the limit \eqref{eq:ab-sing}.
We then need to construct a new source $u_T(k)$ which, in this limit,
satisfies the condition $[1\!-\!u_T(k)]\to0$.  This new source must
include the hard parton recoils.  Actually we need to consider the
recoil only for the contribution to the source associated to the theta
function fixing the observable $\Ko^T$.  To do this we start by
expanding the functional in \eqref{eq:T-sum1}
\begin{equation}
  \label{eq:T-sum2}
\begin{split}
\exp\left\{\int[dk]\,W_{\conf}(k)\,u_T^{(0)}(k)\right\}\>=\>
\prod_{(ab)}\sum_{n_{ab}}\frac{1}{n_{ab}!}
\prod_{i=1}^{n_{ab}} C^{(\conf)}_{ab}
\int [dk_i]\,w^a_{ab}(k_i)\,u_T^{(0)}(k_i)\>.
\end{split}
\end{equation}
Consider a given term in this expansion. For any given ordered pair
$(ab)$ there are $n_{ab}$ soft gluons which are associated to the
distribution $w^a_{ab}$ and are then emitted in the region $F_{ab}$.
For this term in the expansion \eqref{eq:T-sum2}, the corresponding
theta function fixing the observable $\Ko^T$ gives
\begin{equation}
  \label{eq:decoKoT}
  \Ko=\sum_a|q_{ax}|+{\sum_{i}}\,|k_{ix}|=
\sum_a|q_{ax}|+ \sum_{(ab)}\left\{\sum_{i\in F_{ab}} |k_{ix}|\right\}.
\end{equation}
Recall that the $x$-components are the same in both the laboratory and
the $ab$-dipole c.m. frames.  We then introduce the subtractions
\begin{equation}
  \label{eq:decoKoT1}
\begin{split}  
&\Ko\simeq 
\sum_a|\bar q_{ax}|\! +\!  \sum_{(ab)}\left\{\sum_{i\in F_{ab}}
\left(|k_{ix}|\! -\! \al'_i |\bar q_{ax}|\right)\right\},
\\&
\bar q_{ax}\equiv
\left(1\! +\! \sum_{b\ne a}\sum_{i\in F_{ab}}\al'_i\right)\>q_{ax}\,,
\end{split}
\end{equation}
($\al'_i$ is defined in the $ab$-dipole c.m.).  Here we have neglected
corrections of second order in the soft parameters.

Now we reconstruct the expansion \eqref{eq:T-sum2} by using
\eqref{eq:decoKoT1} instead of \eqref{eq:decoKoT}. While
\eqref{eq:decoKoT} gives
\eqref{eq:T-sum}, using \eqref{eq:decoKoT1} we obtain
\begin{equation}
  \label{eq:T-sum3}
\begin{split}
\sum_n\frac{1}{n!}\int dH_n^T\,\prod_i[dk_i]\,M_{n,\conf}^2
=\int&\frac{d\nu\,e^{\nu \Ko} }{2\pi i \nu}
\left\{\nu^4 \frac{d\gam dq_{1y}}{2\pi} 
\prod_{a=1}^3
\frac{d\be_{a}d\bar q_{ax}}{2\pi}\, V(\bar q)\right\}\\
&\cdot \prod_{(ab)}\exp\left\{C^{(\conf)}_{ab}\int\,
[dk]\,w^a_{ab}(k)\,u_{ab}(k)\right\}\>,
\end{split}
\end{equation}
where $u_{ab}(k)$ is a source associated to the ordered pair $(ab)$
given by
\begin{equation}
  \label{eq:uabT}
u_{ab}(k)=u_T^{(0)}(k)\cdot e^{\nu\al'|\bar q_{ax}|}
=e^{-\nu(|k_x|-\al'|\bar q_{ax}|)}\cdot \hat u_T(k)\>,
\end{equation}
where $\hat u_T(k)$ is the part of the source depending on
the Fourier variables $\be_a,\gam$ (see \eqref{eq:u0ell}).
The factor $e^{-\nu(|k_x|-\al'|\bar q_{ax}|)}$ , which is the only one
relevant for the NP corrections, is equal to one at the singular
point \eqref{eq:ab-sing}. We have then that \eqref{eq:uabT} is the
form of the source we are searching for.

The Jacobian  for changing integration variables from $q_{ax}$ to
$\bar q_{ax}$ is given by
\begin{equation}
  \label{eq:JacT}
  J(\al)=\prod_a\left(1+\sum_{b\ne a} \sum_{i\in F_{ab}}\al'_i\right)^{-1}
\simeq \prod_{(ab)} \prod_{i\in F_{ab}}(1\!-\!\al'_i) \>.
\end{equation}
This can be taken into account as a correction to the soft
distribution $w^a_{ab}(k)\to (1-\al')w^a_{ab}(k)$. This correction
corresponds to including the hard part of the splitting function, see
\cite{DLMS}. Since in \eqref{eq:T-sum3} the rescaled momentum $\bar
q_{ax}$ is just an integration variable, in the following we shall
rename it $q_{ax}$.

\subsection{NP radiator for the $\Ko^T$ distribution \label{App:NPRadT}}
The radiator for the $\Ko^T$-distribution, given in
terms of the dipole components \eqref{eq:decow}, reads
\begin{equation}
  \label{eq:R1}
\cR_{\conf}\!=\!\sum_{a<b}C_{ab}^{(\conf)}\,r_{ab}\>,
\quad r_{ab}\!=\!r^a_{ab}\!+\!r^b_{ba}\>,
\quad r^a_{ab}\equiv 
\int\frac{d^3k}{\pi\om}w^a_{ab}(k)[1\!-\!u_{ab}(k)]\,,
\end{equation}
where the integration region is restricted to the $F_{ab}$ region.  

The distribution for the emission of two soft partons in the three-jet
event (see \cite{2soft} and \cite{acopt}) can be represented as the
sum of the two soft parton emission from the three dipoles.  This fact
can be used to obtain the two-loop PT radiator as
sum of the two-loop radiator for each dipole. At this point we can use
all the results of the two-jet analysis.
At PT level this gives rise simply to the identification of the
running coupling scale for each dipole radiator, the invariant dipole
transverse momentum, and the scheme. 
For the analysis of the NP corrections we follow the two-jet case
method recalled above. Introducing the coupling according to the
dispersive method \cite{DMW} we write the two-loop dipole radiator as
a sum of three pieces, the naive, the inclusive and the non-inclusive
pieces.  The naive piece of the $ab$-dipole is given, as in
\eqref{eq:2R}, by
\begin{equation}
  \label{eq:rab-naive}
r^a_{ab}=\int dm^2\aef(m^2)\frac{-d}{dm^2}
\int \frac{d^2\ka}{\pi (\ka^2+m^2)} 
\int_{\frac{\sqrt{\ka^2+m^2}}{Q_{ab}}}^{1}\frac{d\al'}{\al'}\>
[1-u_{ab}(k)]\>,
\end{equation}
with $\aef(m^2)$ the effective coupling discussed in \cite{DMW}.  The
integration region is restricted to the forward region $F_{ab}$
defined by $\al'> k'_t/Q_{ab}$. This has been replaced by
$\al'>\sqrt{\ka^2+m^2}/Q_{ab}$, since, at the lower limit of $\al'$
(large angle emission) the correction is of second order in the
NP-parameter (see \cite{broad}).

As discussed in the Appendix \ref{App:2jet}, the leading NP correction
coming from \eqref{eq:rab-naive} is obtained as follows:
\begin{itemize}
\item in the integrand we take the linear contribution in $\ka,m\sim
  \LQCD\ll|q_{ax}|$.  Therefore, for the massless gluon source
  \eqref{eq:uabT}, we take the first term in the expansion
\begin{equation}
  \label{eq:ulim0}
[1- u_{ab}(k)]=\nu(|k_x|- \al'|q_{ax}|)+\dots
\qquad k_x=\ka\cos\phi +\al'q_{ax}\>,
\end{equation}
where $\phi$ is the azimuthal angle of $\vka$. The neglected terms
contribute to the higher order power corrections;
\item for a gluon with mass $m$ we make the replacement
  \eqref{eq:mass}. This ensures a simple treatment of the other
  (inclusive and non-inclusive) pieces of the two-loop radiator;
\item we take the NP piece $\daef(m^2)$ of the effective coupling. 
\end{itemize}
The leading NP correction to the naive piece is then given by
\begin{equation}
  \label{eq:npab'}
\begin{split}
&\de  r^a_{ab}=\int dm^2\frac{\daef(m^2)}{\pi}\frac{-d}{dm^2}
\int \frac{d\ka^2}{\ka^2+m^2}\>\Om_{ab}(\ka^2+m^2)\>,\\
&\Om_{ab}(\ka^2+m^2)=\nu
\int_{-\pi}^{\pi}\frac{d\phi}{2\pi}
\int_{\frac{\sqrt{\ka^2+m^2}}{Q_{ab}}}^{1}\frac{d\al'}{\al'}
\left( |\sqrt{\ka^2+m^2}\,\cos\phi+\al'q_{ax}|-\al'|q_{ax}|\right).
\end{split}
\end{equation}
To evaluate this integral we introduce the rescaled variable 
$$
v=\frac{\al'|q_{ax}|}{\sqrt{\ka^2+m^2}}
$$
and find
\begin{equation}
\label{eq:Om'}
\Om_{ab}(\ka^2+m^2)=\nu \sqrt{\ka^2+m^2}\>\int_{-\pi}^{\pi} \frac{d\phi}{2\pi}
\int_{\frac{|q_{ax}|}{Q_{ab}}}^{1}
\frac{dv}{v}\left(\,|\cos\phi-v|-v\,\right)\,.
\end{equation}
Notice that at the upper limit $\al'=1$ one has
$v=\frac{|q_{ax}|}{\sqrt{\ka^2+m^2}}$ which tends to infinity in the
limit we are considering.  On the other hand the contribution with
$v>1$ vanishes due to symmetry of $\phi$-integration.  Setting
$q_{1x}\to 0$ at the $v$-lower limit, we find
\begin{equation}
  \label{eq:xx}
\Om_{ab}(\ka^2+m^2)=
\nu\,c_{\Ko}\,\sqrt{\ka^2+m^2}\> \ln \frac{\zeta Q_{ab}}{|q_{ax}|}\>,
\quad c_{\Ko}=\frac{2}{\pi}\>,
\end{equation}
with $c_{\Ko}$ the characteristic number for the $\Ko$ observable and
$\zeta$, which takes into account the small-angle contribution, is
given by
\begin{equation}
\label{eq:zeta}
\begin{split}
\ln \zeta=\frac{\pi}{2}
\int_{-\pi}^{\pi} \frac{d\phi}{2\pi}
\int_{0}^{1}
\frac{dv}{v}\big\{|\cos\phi-v|-v-|cos\phi|\big\}=\ln2 - 2 \>.
\end{split}
\end{equation}
We get
\begin{equation}
  \label{eq:rNP}
\de r^a_{ab}(q_{ax})=\nu\cp\> \ln \frac{\zeta Q_{ab}}{|q_{ax}|}\>, 
\qquad 
\cp = c_{\Ko}\,\int \frac{dm^2}{m^2}\frac{\daef(m)}{\pi}\>m\>.
\end{equation}
For the ``inclusive'' and ''non-inclusive'' pieces the analysis is the
standard one and we have that the full NP correction is given by
\eqref{eq:rNP} by rescaling the NP-parameter by the Milan factor $\cM$.

The NP parameter $\cp$ has dimension of a mass, and this implies that
the NP correction is proportional to $1/Q$.
It can be expressed in terms of the integral of the running coupling over 
the infrared region
\begin{equation}
  \label{eq:S-al0}
\al_0(\mu_I)=\int_0^{\mu_I}\frac{dk}{\mu_I}\as(k)\>.
\end{equation}
This parameter $\al_0(\mu_I)$ is the same entering the two-jet shape
variables $1-T$, $C$, $B$ and $M^2/Q^2$.  After merging PT and NP
contributions to the observable in a renormalon free manner, one has
that the distribution is independent of $\mu_I$ and one obtains 
\begin{equation}
  \label{eq:cp}
 \cp\> \equiv\>  c_{\Ko}\,\cM \frac{4}{\pi^2}\mu_I
\left\{ \alpha_0(\mu_I)- \bar{\as}
  -\beta_0\frac{\bar{\as}^2}{2\pi}\left(\ln\frac{Q}{\mu_I} 
+\frac{K}{\be_0}+1\right) \right\}\>,  
\quad \bar{\as}\equiv \al_{\MSbar}(Q)\>.
\end{equation}
The term proportional to $K$ accounts for mismatch between the
$\MSbar$ and the physical scheme \cite{CMW} and is given by
\begin{equation}
   K\equiv
  C_A\left(\frac{67}{18}-\frac{\pi^2}{6}\right)-\frac{5}{9}n_f \>,
  \qquad \beta_0=\frac{11N_c}{3}-\frac{2n_f}{3}\>.
\end{equation}
It may be useful to connect this parameter entering the shift of the
$\Ko$ distributions to the analogous one for the thrust distribution.
Defining $\tau=1\!-\!T$, we have
\begin{equation}
  \label{eq:thrust}
  \frac{d\sigma}{d\tau}(\tau)=
\frac{d\sigma^{\PT}}{d\tau}(\tau\!-\!\Delta_{\tau})\>,
\qquad   \Delta_{\tau}=C_F\,\frac{c_{\tau}\,\cp}{c_{\Ko}}\>,
\end{equation}
where $C_F$ enters due to the fact that the two-jet system is made of
a quark-antiquark pair, $c_{\tau}=2$ and $c_{\Ko}=2/\pi$ are the
characteristic numbers associated to the phase space integration of
the relative observables.

We can now reconstruct the NP correction to the full radiator in
\eqref{eq:R1} and we find
\begin{equation}
  \label{eq:fineT}
\begin{split}
&  \de\cR_3(q)=\nu\cp\left(
C_F\ln\frac{\zeta Q_{12}}{|q_{1x}|}+
C_F\ln\frac{\zeta Q_{12}}{|q_{2x}|}+
C_A\ln\frac{\zeta Q_{23}Q_{13}}{Q_{12}|q_{3x}|}\right)\,,\\
&  \de\cR_2(q)=\nu\cp\left(
C_F\ln\frac{\zeta Q_{13}}{|q_{1x}|}+
C_F\ln\frac{\zeta Q_{13}}{|q_{3x}|}+
C_A\ln\frac{\zeta Q_{23}Q_{12}}{Q_{13}|q_{2x}|}\right)\,,\\
&  \de\cR_1(q)=\nu\cp\left(
C_F\ln\frac{\zeta Q_{23}}{|q_{3x}|}+
C_F\ln\frac{\zeta Q_{23}}{|q_{2x}|}+
C_A\ln\frac{\zeta Q_{12}Q_{13}}{Q_{23}|q_{1x}|}\right)\,.
\end{split}
\end{equation}
We find then that the NP scales are given by the hard scales $Q_a$ in
\eqref{eq:Qa} rescaled by a factor $\zeta^{\NP}=\zeta$ (see \eqref{eq:QNP}).

\subsection{NP corrections to the $\Ko^R$ distribution \label{App:NPRadR}}
We proceed as in the previous case. Here the simplification is that we
should be careful only with the collinear singularity with respect to
$p_1$ which is contained only in the distributions $w^1_{1b}(k)$. 

First we consider the modification of the sources.
We must include the recoil only in the source $u_{1b}(k)$ associated
to the distribution $w^1_{1b}(k)$. We have
\begin{equation}
  \label{eq:u1bR}
  u_{1b}(k)=\Theta_L(k)+\Theta_R(k)\,e^{-\nu(|k_x|-\al'|q_{1x}|)}\,
e^{i\be k_x}\>.
\end{equation}
For the other sources $u_{ab}(k)$, associated to the distribution
$w^a_{ab}(k)$ with $a\ne1$, we do not need modifications:
$u_{ab}(k)=u^{(0)}_R(k)$.  The modification in \eqref{eq:u1bR} is
accompanied by a change of variable from $q_{1x}$ to $\bar q_{1x}$
obtained by a rescaling factor similar to \eqref{eq:decoKoT1}. However
since $\bar q_{1x}$ is an integration variable we shall rename it
$q_{1x}$.

Then we take the linear contribution for the massive gluon source
\begin{equation}
  \label{eq:ulimR}
\begin{split}
& [1-u_{1b}(k)] \quad \Rightarrow\quad
\nu\left(|\sqrt{\ka^2+m^2}\>\cos \phi+\al' q_{1x}|-
  \al'|q_{1x}|\right)
\,\Theta_R(k)\,,\\
& [1-u_{ab}(k)] \quad \Rightarrow\quad\nu\sqrt{\ka^2+m^2}|\cos
\phi|\,\Theta_R(k)\>, \qquad a\ne1\>.
\end{split}
\end{equation}
Since the distribution has a singularity for the soft gluon collinear
to $p_1$, the NP radiator has a logarithmic behaviour in $q_{1x}$. We
have
\begin{equation}
  \label{eq:inizioR}
\de\cR_{\conf}(q) = \nu\cp C_1^{(\conf)}
\ln\frac{Q^{\NP}_{1}}{|q_{1x}|}\,.
\end{equation}
The $\ln |q_{1x}|$ part is contained in $r^1_{12}$ and $r^1_{13}$
since the corresponding distributions are collinear singular in the
integration region.  The remaining pieces contribute only to set the
NP scale.  To obtain the NP hard scale $Q^{\NP}_{1}$ we need to
analyze the individual pieces.

To see the singular behaviour in \eqref{eq:inizioR} consider the NP
correction to $r^1_{1b}$.  At two loop, the naive term of $r^1_{1b}$
is given in terms of the effective coupling as follows
\begin{equation}
  \label{eq:np1b}
r^1_{1b}=\int dm^2\aef(m^2)\frac{-d}{dm^2}
\int \frac{d^2\ka}{\pi (\ka^2+m^2)} 
\int_{\frac{\sqrt{\ka^2+m^2}}{Q_{1b}}}^{1}\frac{d\al'}{\al'}\>
[1-u_{1b}(k)]\>.
\end{equation}
Taking the linear part of the source and NP part of the effective
coupling, we have
\begin{equation}
  \label{eq:npab2'}
\begin{split}
&\de  r^1_{1b}=\int dm^2\daef(m^2)\frac{-d}{dm^2}
\int \frac{d\ka^2}{\ka^2+m^2}\>\Om_{1b}(\ka^2+m^2)\>,\\
&
\Om_{1b}(\ka^2+m^2)\>=
\>\nu \sqrt{\ka^2+m^2}\>\int_{-\pi}^{\pi} \frac{d\phi}{2\pi}
\int_{\frac{|q_{ax}|}{Q_{1a}}}^{1}
\frac{dv}{v}\left(\,|\cos\phi-v|-v\,\right)\Theta_R(k)\>,
\end{split}
\end{equation}
where we need to express the condition in $\Theta_R$ in terms of the
variables in the $1b$-dipole c.m. system (see \eqref{eq:Sab}).

\subsection{Recoil integration over the NP radiator}
For the $\Ko^T$ case we have
\begin{equation}
  \label{eq:T-INP}
\begin{split}
&\frac{\nu^4}{8} \int dq_{1y} \prod_adq_{ax}\> V^T(\gam\be q)
\>\prod_a\left(\frac{|q_{ax}|}{Q_a^{\NP}}\right)^{\nu C_a\cp}\\
& \simeq I(\be,\gam)\left\{1-\nu\cp \sum_a C_{a}\,
\left[\> \ln( \bnu Q_{a}^{\NP})\>+\>\tchi_a(\be,\gam)\>\right]\right\},
  \end{split}
\end{equation}
with $V^T$ and $I(\be,\gam)$ given in \eqref{eq:VT} and
\eqref{eq:T-I0} respectively, $\tchi_2\equiv\chi(\be_2)$,
$\tchi_3\equiv\chi(\be_3)$ (see \eqref{eq:chi}) and
\begin{equation}
  \label{eq:tchi}
\begin{split}
 I(\be,\gam)\,\tchi_1\equiv
\frac{1}{(1+\be^2_2)(1+\be^2_3)}
\left(\frac{\chi(\be_{12})}{(1+\be^2_{12})(-i\gam+\eps)}+
      \frac{\chi(\be_{13})}{(1+\be^2_{13})( i\gam+\eps)}\right).
\end{split}
\end{equation}
For the $\Ko^R$ case we have 
\begin{equation}
  \label{eq:R-INP}
\begin{split}
\frac{\nu}{2} \int_{-\infty}^{\infty} dq_{1x}\, V^R(\be,q_{1x})
\left(\frac{|q_{1x}|}{Q_{1}^{\NP}}\right)^{\nu C_1\cp}\!\!
\simeq\>
\frac{1\!-\!\nu\,\cp\,C_1\,
\left[\, \ln( \bnu Q_{1}^{\NP})\!+\!\chi(\be)\right]}{1+\be^2}\>,
\end{split}
\end{equation}
where $V^R$ is given in \eqref{eq:VR}.  Both in \eqref{eq:T-INP} and
\eqref{eq:R-INP}, corrections are of order $(\cp)^2$.

\section{The NP functions $\tf_a$ \label{App:tfa}}
\subsection{$\tf_1$ \label{App:tf1}}
To compute $\tf_1$ we first split the function $\chi(\be)$ in the two 
pieces in \eqref{eq:chi} and evaluate the two contributions separately.
We have $\tf_1(\nu)=\tf'_1(\nu)+\tf''_1(\nu)$ where 
\begin{equation}
\label{eq:tf1app}
\begin{split}  
&\tf'_1(\nu)=\int_{-\infty}^{\infty}\prod_{a=2,3} 
\frac{d\be_a\,e^{-R_a\left(\bnu\sqrt{1+\be_a^2}\right)}}{\pi(1+\be_a^2)}
\int_{-\infty}^\infty 
\frac{d\be\,\chi'(\be)}{\pi(1+\be^2)}\,\cS(\nu,\be,\be_{23})\>,\\
&\tf''_1(\nu)=\int_{-\infty}^{\infty}\prod_{a=2,3} 
\frac{d\be_a\,e^{-R_a\left(\bnu\sqrt{1+\be_a^2}\right)}}{\pi(1+\be_a^2)}
\int_{0}^\infty \frac{\be d\be}{1+\be^2}\,\cS(\nu,\be,\be_{23})\>.
\end{split}
\end{equation}
The evaluation $\tf_1'$ is straightforward since all three
$\be$-integrals are rapidly convergent.  For large $\nu$ we can
factorize the three radiators and the remaining piece is a function
of the SL variable $r'(\nu,Q)$
\begin{equation}
  \label{eq:tf1'}
  \tf'_1(\nu)=\prod_{a=1}^3e^{-R_a(\bnu)}\cdot\cC'_1(\nu)\>,
\qquad \cC'_1(\nu) = R'_1(\bnu)A_1\left(\{R_a'(\bnu)\}\right)\>.
\end{equation}
For $\nu=0$, we set to zero the radiator and, due to the integration
property of $\chi'$ in \eqref{eq:chi}, we have $\tf'_1(0)=0$.

To evaluate $\tf_1''$ we observe that the $\be_2$- and
$\be_3$-integrals are rapidly convergent, so that the characteristic
values of these two variables are $\be_2,\be_3=\cO1$.  On the contrary
the $\be$-integral is logarithmic so that the characteristic values of
$\be$ are large.  
From \eqref{eq:T-Rad1-ri}, \eqref{eq:bmu} and \eqref{eq:Blim}, in the
region
\begin{equation}
  \label{eq:tf1-ps}
\be_2,\,\be_3\ll\be\simeq \be_{T}\>,  
\end{equation}
we have 
\begin{equation}
  \label{eq:tf1-Radlim}
  \bR_1^{(r)}(\bnu,\be,\bbe,0)\simeq R_1(\bnu\be)\,\qquad   
  \bR_1^{(i)}(\bnu,\be,\bbe,\gam)\to 0\,,
\end{equation}
(see \eqref{eq:T-Rad1-ri} and \eqref{eq:Blim}). Therefore, in $\cS$
one can factorize the $\#1$ radiator and the remaining factor is a SL
function
\begin{equation}
  \label{eq:tf1-cS}
\cS(\nu,\be,\be_{23})=e^{-R_1(\rho)}
\cdot\left\{\,1+ R_1'(\bnu)\,B_1
\left(\be,\be_{23},R'_1(\bnu)\right)\right\},
\quad \rho\equiv\bnu\sqrt{1+\be^2}\>,
\end{equation}
with $B_1$ vanishing in region \eqref{eq:tf1-ps}.  The first
contribution gives the leading contribution to $\tf_1$. We can
factorize the $\be_2$- and $\be_3$-integrals and, changing variable
from $\be$ to $\rho$ we obtain
\begin{equation}
  \label{eq:tf10}
\tf_1(\nu)\> \simeq \prod_{a=2,3}\int_{-\infty}^{\infty}  
\frac{d\be_a\,e^{-R_a\left(\bnu\sqrt{1+\be_a^2}\right)}}{\pi(1+\be_a^2)}
\int_{\bnu}^{\infty}\frac{d\rho\,e^{-R_1(\rho)}}{\rho}
=
\prod_{a=1}^3e^{-R_a(\bnu)}\,\cF_2(\nu)\,\cF_3(\nu)\,E_1(\bnu)\>,
\end{equation}
where we have factorized the three radiators, the functions $\cF_a$
are the single jet functions defined in \eqref{eq:cFa} and the
function $E_1$ is introduced in \eqref{eq:E1} and discussed in detail
in Appendix \ref{App:E1}.

Consider the second term contribution in \eqref{eq:tf1-cS} to
$\tf_1''$. It will be denoted by $\tf'''_1$. Since $B_1$ vanishes for
large $\be$, all three $\be$-integrals are rapidly convergent. As for
$\tf_1'$, we can factorize the three radiators and the remaining
integrals give a SL function
\begin{equation}
  \label{eq:tf1'''}
    \tf'''_1(\nu)=\prod_{a=1}^3e^{-R_a(\bnu)}\cdot\cC''_1(\nu)\>,
\qquad \cC''_1(\nu)=R'_1(\bnu) A_1'\left(\{R_a'(\bnu)\}\right)\>.
\end{equation}
Here again we have that $\cC''_1$ vanishes for $R_1'\to0$ and $\nu\to0$.
Assembling all pieces \eqref{eq:tf1'}, \eqref{eq:tf10}, and
\eqref{eq:tf1'''}, the function $\tf_1(\nu)$ is given by
\begin{equation}
  \label{eq:tf1-fine}
  \tf_1(\nu)=\prod_{a=1}^3e^{-R_a(\bnu)}\,\left\{
\cF_2(\nu)\,\cF_3(\nu)\,E_1(\bnu)\>+\> \cC_1(\nu)\right\}\>.
\end{equation}
Here $ \cC_1(\nu)=\cC'_1(\nu)+\cC''_1(\nu)$ is a SL function, which
vanishes for $R_1'\to0$ and for $\nu\to0$. .

To conclude we compute the first loop contribution proportional to $R'_1$
to $\cS$. We have 
\begin{equation}
  \label{eq:cS-onel}
\cS(\nu,\be,\be_{23})\, e^{R_1(\rho)}\>=\>
1+\half R'_1(\bnu)\ln \sqrt{\frac{1+\be^2}{1+\bbe^2}}
+\frac{2}{\pi}\int_0^{\infty}\frac{d\gam}{\gam}\>\Im
\bR_1(\bnu,\be,\bbe,\gam)
+\cdots
\end{equation}
The first two terms are the $\gam=0$ contribution. The last term 
has been evaluated in \cite{acopt} and one finds
\begin{equation}
\label{eq:Im-onel}
 \frac{2}{\pi}\int_0^\infty \frac{d\gam}{\gam}\, 
\Im\,\bR_1(\bnu,\be,\bbe,\gam)=
\frac{R_1'(\bnu)}{2}\cdot \ln\frac{\sqrt{1+\be^2}}{\sqrt{1+\bbe^2}}\>.
\end{equation}
Thus the one loop contribution is given by twice the second term.
It vanishes for $\be\to\infty$.

\subsection{$\tf_2$ \label{App:tf2}}
As before, to evaluate $\tf_2$ we first split the function
$\chi(\be_2)$ in the two pieces in \eqref{eq:chi} and compute the
two contributions.  We have $\tf_2(\nu)=\tf'_2(\nu)+\tf''_2(\nu)$,
where
\begin{equation}
\label{eq:tf2app}
\begin{split}  
&\tf'_2(\nu)=\int_{-\infty}^{\infty}\prod_{a=2,3} 
\frac{d\be_a\,e^{-R_a\left(\bnu\sqrt{1+\be^2_a}\right)}}{\pi(1+\be_a^2)}
 \int_{-\infty}^\infty 
\frac{d\be\,\cS(\nu,\be,\be_{23})}{\pi(1+\be^2)}\, \chi'(\be_2)\>,\\
&\tf''_2(\nu)=\int_{0}^{\infty}
\frac{\be_2d\be_2\,e^{-R_2\left(\bnu\sqrt{1+\be^2_2}\right)}}{1+\be_2^2}
\int_{-\infty}^{\infty}
\frac{d\be_3\,e^{-R_3\left(\bnu\sqrt{1+\be^2_3}\right)}}{\pi(1+\be_3^2)}
\int_{-\infty}^\infty 
\frac{d\be\,\cS(\nu,\be,\be_{23})}{\pi(1+\be^2)}\>.
\end{split}
\end{equation}
The analysis of $\tf_2'$ is similar to the one of $\tf_1'$ performed
before and we get 
\begin{equation}
  \label{eq:tf2'}
  \tf'_2(\nu)=\prod_{a=1}^3e^{-R_a(\bnu)}\cdot \cC_2'(\nu)\>,
\qquad \cC'_2(\nu)=R'_2(\bnu) A_2\left(\{R_a'(\bnu)\}\right)\>,
\end{equation}
where $\cC'_2(\nu)$ is a SL function vanishing for $R'_2\to0$ and
$\nu\to0$.

To evaluate $\tf_2''$ we observe that the $\be_3$- and $\be$-integrals
are rapidly convergent, so that the characteristic values of these two
variables are $\be,\be_3=\cO1$.  On the contrary the $\be_2$-integral
is logarithmic so that the characteristic values of $\be_2$ are large.
Therefore, to evaluate $\tf_2''$ we need to obtain
$\cS(\nu,\be,\be_{23})$ in the region
\begin{equation}
  \label{eq:tf2-ps}
\be,\,\be_3\ll\be_2 \>.
\end{equation}
We show that here one has
\begin{equation}
  \label{eq:app-cS2}
  \cS(\nu,\be,\be_{23})=e^{-R_1\left(\bnu\sqrt{1+\be_2^2}\right)}\cdot
\left\{1+R_1'(\bnu)\,B_2(\nu,\be,\be_{23})\right\},
\end{equation}
where $B_2$ is a SL function of $\nu$ and rapidly vanishes in the region
\eqref{eq:tf2-ps}.

To obtain this result we observe that in this region we must consider
also the second term of $\cS$ in \eqref{eq:cS} with the
$\gam$-integral and the imaginary part of $\bR_1$. This integral is
logarithmic, as the $\be_2$-integral, so that the characteristic
values of $\gam$ are large compared to $\be,\be_3$.  On the other
hand, from \eqref{eq:T-Rad1-ri}, \eqref{eq:bmu} and \eqref{eq:Blim},
we have $\bR_1^{(i)}(\bnu,\be,\bbe,\gam)\to 0$ for $\be_T\ll\gam$. We
conclude that the region

\begin{equation}
  \label{eq:leading}
\be,\,\be_3\ll\gam\ll\be_T\simeq\be_2\simeq\be_{23}\>,
\end{equation}
gives the leading logarithmic approximation of $\tf''_2$.  We start
from this approximation which we shall later improve to account for
the first subleading correction $\cO{R_1'}$.  In this region, from
\eqref{eq:T-Rad1-ri}, we have
\begin{equation}
\label{eq:ap1}
\begin{split}
  &\bR^{(r)}(\bnu,\be,\be_T,\gam)\simeq
  \half R_1(\bnu\gam)+\half R_1(\bnu\be_2)\>,\\
  &\bR^{(i)}(\bnu,\be,\be_T,\gam)\simeq -\frac{\pi}{4}
  R'_1(\bnu\gam)\>,
\end{split}
\end{equation}
and, for the two terms of $\cS$ in \eqref{eq:cS}, we have
\begin{equation}
\label{eq:ap2}
\begin{split}
e^{-\bR_1(\bnu,\be,\bbe,0)}\>&\simeq\> 
e^{-\half R_1(\bnu\be_2)}\,e^{-\half R_1\left(\bnu\sqrt{1+\be^2}\right)}\>,\\
\int_0^{\infty}\frac{d\gam}{\gam}\>\Im\,
e^{-\bR_1(\bnu,\be,\bbe,\gam)}\>&\simeq\> 
e^{-\half R_1(\bnu\be_2)}\>
\int_{\gam_m}^{\gam_M}\frac{d\gam}{\gam}\>
e^{-\half R_1(\bnu\gam)}\>\sin\left(\frac{\pi}{4} R'_1(\bnu\gam)\right),
\end{split}
\end{equation}
where $\gam_M=\cO{\be_2}$ and $\gam_m=\cO{\be}$.  
To evaluate the contribution from the imaginary part we first consider
\begin{equation}
  \label{eq:ap3}
\sin\left(\frac{\pi}{4} R'_1(\bnu\gam)\right)\>=
\>\frac{\pi}{4} R'_1(\bnu\gam)+\cdots
\end{equation}
The terms in the dots can be neglected since they contribute to the
integral with terms of order $\as$.  To show this consider the
contribution to the $\gam$-integral from the first term which is of
order $(\as\ln\gam)^3$ for large $\gam$.  Since the $\gam$ integration
is logarithmic and the radiator is of order $\as\ln^2\gam$, the result
of the integral is of order $\as$.

Taking the first term in \eqref{eq:ap3}, we find the leading
logarithmic contribution
\begin{equation}
  \label{eq:f2log}
\begin{split}
-\frac{2}{\pi}\int_0^{\infty}\frac{d\gam}{\gam}\>\Im\,
e^{-\bR_1(\bnu,\be,\bbe,\gam)}
&\simeq\>e^{-\half R_1(\bnu\be_2)} \int_{\gam_m}^{\gam_M}
d\gam\>\frac{d}{d\gam}\left\{ e^{-\half R_1(\bnu\gam)}\right\}\\
& \simeq\> e^{-R_1(\bnu\be_2)}\>-\>
e^{-\half R_1(\bnu\be_2)-\half R_1(\bnu\gam_m)}\>.
\end{split}
\end{equation}
Here we used $\gam_M\sim\be_2$.  

In order to keep under control the subleading $\cO{r'}$ correction we
need to fix precisely the lower integration limit $\gam_m$ in
\eqref{eq:f2log}.  To do this it suffices to calculate the {\em
  exact}\/ $\gam$-integral in the first order in $r'$ given in
\eqref{eq:Im-onel}.
Combining \eqref{eq:f2log} and \eqref{eq:Im-onel} we have the result
\begin{equation}
-\frac{2}{\pi}\int_0^\infty \frac{d\gam}{\gam}\> 
\Im\, e^{-\bR_1(\bnu,\be,\bbe,\gam)}
\simeq e^{-R_1\left(\bnu\sqrt{1+\be_2^2}\right)}\>-\>
e^{-\half R_1\left(\bnu\sqrt{1+\be^2_2}\right)
   -\half R_1\left(\bnu\sqrt{1+\be^2  }\right)}\>.
\end{equation}
Summing this with the real part contribution in \eqref{eq:ap2} we have
the cancellation of the leading term $e^{-\half
  R_1(\bnu\be_2)}e^{-\half R_1(\bnu\sqrt{1+\be^2})}$ and
we obtain
\begin{equation}
  \label{eq:appcS}
  \cS(\nu,\be,\be_{23})\simeq e^{-R_1(\rho)}\>, 
\qquad \rho\equiv \bnu\sqrt{1+\be_{2}^2}\>.
\end{equation}
This result shows that if one factorizes in $\cS$ the radiator
$e^{-R_1(\rho)}$, the remaining piece is a SL function and one obtains
the final result in \eqref{eq:cS2}.

The leading piece to $\tf_2$ is obtained from \eqref{eq:appcS}
and is given by
\begin{equation}
  \label{eq:tf20}
  \tf_2(\nu)\simeq
\int\frac{d\be_3e^{-R_3(\bnu\sqrt{1+\be_3^2})}}{\pi(1+\be_3^2)}\,
\int_{\bnu}^{\infty}\frac{d\rho}{\rho}\,
e^{-R_1(\rho)-R_2(\rho)}=
\prod_{a=1}^3e^{-R_a(\bnu)}\cdot\cF_3(\nu)\,E_2(\bnu)\>,
\end{equation}
with $E_2$ given in \eqref{eq:E2} and $\cF_3$ given by \eqref{eq:cFa}.

The second term in \eqref{eq:app-cS2} (with $B_2$) gives, after
factorizing the radiators, a SL function since all three
$\be$-integrals are rapidly convergent. One gets
\begin{equation}
  \label{eq:tf2'''}
  \tf'''_2(\nu)=\prod_{a=1}^3e^{-R_a(\bnu)}\cdot\cC''_2(\nu)\>,\qquad
  \cC''_2(\nu)=R'_1(\nu)\,A'_2(\nu)\>,
\end{equation}
with $A'_2$ a SL function. 

Putting together the various pieces, \eqref{eq:tf2'}, \eqref{eq:tf20}
and \eqref{eq:tf2'''}, we get the final result
\begin{equation}
  \label{eq:tf2-fine}
  \tf_2(\nu)=\prod_{a=1}^3e^{-R_a(\bnu)}\cdot\left\{
\cF_3(\nu)E_{2}(\nu)+ \cC_2(\nu)\right\}\>,
\end{equation}
where $\cC_2(\nu)=R_2'(\bnu)A_2(\nu)+R_1'(\bnu)A'_2(\nu)$ is a SL
function which vanishes for $r'\to0$ and $\nu\to0$.

\section{The $E$ functions \label{App:E}}
For completeness we recall here the results of an Appendix of
Ref.~\cite{broad} and extend them to the present case.

\subsection{$E_1$}\label{App:E1}
We consider first the function $E_1(\bnu)$. 
For the function $E_1(\bnu)$ we have the expansion
\begin{equation}
\label{eq:e1}
\begin{split}
E_1(\bnu)&\equiv \int_{\bnu}^{\infty}\frac{d\rho}{\rho}\,
e^{-(R_1(\rho)-R_1(\bnu))}\\&=
\sqrt{\frac{\pi}{2R_1''}}\>N(t) -\frac{R_1'''}{3(R_1'')^2}\>X(t)
+\cO{\as}\>,\qquad t\equiv \frac{R_1'}{\sqrt{2R_1''}}\>,
\end{split}
\end{equation}
with $R_1',R_1'',R_1'''$ the first, second and third logarithmic
derivatives of $R_1$ evaluated at $\bnu$ and
\begin{equation}\label{eq:N-X}
\begin{split}
N(t)\equiv\frac{2}{\sqrt{\pi}}\,\int_t^{\infty}dx\> e^{-x^2+t^2}\>,
\qquad 
X(t)\equiv
-\frac{\sqrt{\pi}}{8}\frac{d^3N(t)}{dt^3}=
\int_0^{\infty}dz\> z\>  e^{-z -2 t \sqrt{z}}\>,
\end{split}
\end{equation}
with the following behaviours
\begin{equation}
\label{eq:expansion}
\begin{split}
& t\gg1: \quad
N(t)= \frac{1}{\sqrt{\pi}t}\left(1-\frac{1}{2t^2}
+\frac{3}{t^4}+\dots\right), 
\quad X(t)= \frac{3}{4 t^4}+ \cO{t^{-6}}\>,\\
& t\ll1:\quad 
N(t)= 1-\frac{2t}{\sqrt{\pi}}+t^2-\frac{4t^3}{3\sqrt{\pi}}+\dots, 
\quad X(t)= 1-\frac{3\sqrt{\pi}}{2}t +4 t^2+\cO{t^3}\>.
\end{split}
\end{equation}
In the region $t=R_1'/\sqrt{2 R_1''}\simeq\sqrt{\as}\ln(\nu Q)\gg1$, we obtain
\begin{equation}
  \label{eq:regione1}
E_1(\bnu)=\frac{1}{R_1'(\nu)}
\left\{1+\cO{\frac{1}{\as\ln^2(\nu Q)}}\right\}\>.
\end{equation}
On the contrary, for $t=R_1'/\sqrt{2  R_1''}\simeq \sqrt{\as}\ln(\nu Q)
\ll 1$,  we find
\begin{equation}
\label{eq:regione2}
 E_1(\bnu) \>=\> 
\frac{\pi}{2\sqrt{C_1\as(Q)}} -\ln(\bnu Q_1^{\PT})  - \frac{\be_0}{6C_1} 
\>+\> \cO{\sqrt{\as}}\>.
\end{equation}
Notice that the term $\cO{\sqrt{\as}}$ implicitly includes all corrections
$\cO{\!\sqrt{\as}}$, 
$\cO{\!\sqrt{\as}\ln\bnu\!}$ 
and $\cO{\sqrt{\as}\ln\bnu^2}$.

\subsection{$E_a$ for $a=2,3$}\label{App:Ea}

For the function $E_a(\bnu)$ we obtain the following result: 
\begin{equation}
\begin{split}
\label{eq:ea}
&E_a(\bnu)\equiv \int_{\bnu}^{\infty}\frac{d\rho}{\rho}\,
e^{-(R_1(\rho)+R_a(\rho)-R_1(\bnu)-R_a(\bnu))}\\
&=\sqrt{\frac{\pi}{2(R_1''+R_a'')}}\>N(t)
- \frac{R_1'''+R_a'''}{3(R_1''+R_a'')^2}\>X(t)+\cO{\as}\>,
\quad t\equiv \frac{R_1'+R_a'}{\sqrt{2(R_1''+R_a'')}}
\end{split}
\end{equation}
where again $R_a',R_a'',R_a'''$ are the first three derivatives of $R_a$
 evaluated at $\bnu$ and the functions $N$ and $X$ are defined in
 \eqref{eq:N-X}. 

As in the previous case, in the region $(R_1'+R_a')/\sqrt{2  (R_1''+R_a'')}
\gg1$, using the expansions in 
\eqref{eq:expansion}, we get
\begin{equation}
  \label{eq:regione1a}
E_a(\bnu)=\frac{1}{R_1'(\nu)
+R_a'(\nu)}
\left\{1+\cO{\frac{1}{\as\ln^2(\nu Q)}}\right\}\>. 
\end{equation}
On the contrary, for $(R_1'+R_a')/\sqrt{2 (R_1''+R_a'')} \ll 1$, we
obtain, up to contributions $\cO{\sqrt{\as}\,}$, 
\begin{equation}
\label{eq:regione2a}
 E_a(\bnu) \simeq 
\frac{\pi}{2\sqrt{(C_1+C_a)\,\as(Q)}} \!-\!
\frac{1}{C_1+C_a}\left(C_1\ln(\bnu Q_1^{\PT})
\!+\!C_a\ln(\bnu Q_a^{\PT})
  \!+\! \frac{\be_0}{6}\right). 
\end{equation}
\newpage

\end{document}